\documentclass[11pt,a4paper]{article}

\usepackage{jcappub,soul}

\title{Scalar Field Dark Matter:  behavior around black holes}

\author{Alejandro Cruz-Osorio,}
\author{F. Siddhartha Guzm\'an,}
\author{Fabio D. Lora-Clavijo}

\affiliation{Instituto de F\'{\i}sica y Matem\'{a}ticas, Universidad
              Michoacana de San Nicol\'as de Hidalgo. Edificio C-3, Cd.
              Universitaria, 58040 Morelia, Michoac\'{a}n,
              M\'{e}xico.}

\emailAdd{alejandro@ifm.umich.mx}
\emailAdd{guzman@ifm.umich.mx}
\emailAdd{fadulora@ifm.umich.mx}

\abstract{
We present the numerical evolution of a massive test scalar 
fields around a Schwarzschild space-time. 
We proceed by using hyperboloidal slices that 
approach future null infinity, which is the  
boundary of scalar fields, and also demand 
the slices to penetrate the event horizon of the black hole.  
This approach allows the scalar field to be accreted by the 
black hole and to escape toward future null infinity. 
We track the evolution of the energy density of the scalar field, 
which determines the rate at which the scalar field is being diluted. 
We find polynomial decay of the energy density of the scalar field, 
and use it  to estimate the rate of dilution of the field in time. 
Our findings imply that the energy density of the scalar field decreases 
even five orders of magnitude in time scales smaller than a year. 
This implies that if a supermassive  black hole is the Schwarzschild solution, 
then scalar field dark matter would be diluted extremely fast.}


\begin{document}

\maketitle


\section{Introduction}

The scalar field dark matter (SFDM) is an alternative to $\Lambda$CDM that predicts flat density profiles in the center of galaxies and fits the abundance of substructure \cite{MatosUrena2000,BernalMatosNunez}, which are two of the problems nowadays of the cold dark matter model. The assumption of this model is that the dark matter is an ultralight scalar field particle, whose mass $m_B$ determines a cut off in the mass power spectrum of structures \cite{MatosUrena2000}. An interesting assumption in such analysis is 
that the scalar field potential is a cosh-like potential, that behaves as 
an exponential at early times and as a free field (quadratic 
potential) at late times. Moreover, it was found that the SFDM enjoys the same 
advantages at cosmic scale as the standard lambda cold dark matter model \cite{MatosUrena2000}.

Initially, within the SFDM model, fully general relativistic stationary solutions were proposed to explain the flatness of rotation curves \cite{GuzmanMatos1999}. Later on, the assumptions relaxed to the 
newtonian limit of such general relativistic models. By the time, the fluid 
dark matter made of scalar fields was proposed as an alternative galactic 
dark matter model \cite{Arbey}. On the other hand, the assumption of time independence was 
also relaxed and scalar field dark matter halos were proposed to be 
gigantic Oscillatons, that is, time dependent fully relativistic scalar 
field solutions to the Einstein-Klein-Gordon system of equations
\cite{galacticolapse,phi2-oscillatons}. On the other hand in order to study late time stages of the structures, the model relaxed to the Newtonian low-energy limit ruled by the
Schr\"odinger-Poisson (SP) system of equations, where the turnaround point and speed of a structure formation was studied \cite{GuzmanUrena2003}; the condition to achieve quick galaxy formation is that the mass of the 
boson is again ultralight ($m_B \sim 10^{-23}$eV) which corresponds well to the cut-off needed in the mass power spectrum \cite{MatosUrena2000}. In \cite{GuzmanUrena2006} it was also shown that the scalar field gravitational collapse tolerates the introduction of a self-interaction 
term in the potential, which makes the model to seem quite like a 
self-gravitating Bose-Condensate whose equilibrium solutions are attractors in time. In \cite{BernalGuzman2006} the non-spherical collapse was studied and shown that again spherical equilibrium solutions are attractor profiles and in \cite{BernalGuzman2006b} it was shown the head-on collision of two structures and interference patterns were found during the collision.

In this paper we explore  the evolution in time of a test scalar field on top of the space-time background of a Schwarzschild black hole, and analyze the time development of the scalar field amplitude and its energy density in order to impose restrictions on the scalar field dark matter model or the system Black Hole plus scalar field dark matter.
An important property of the supermassive black hole plus dark matter system is the asymptotic nature of the space-time. As far as we can tell, black hole candidates are considered to be asymptotically flat, this is the reason why here we consider a Schwarzschild black hole as the model of black hole. Sporadic models like that in \cite{MatosGuzman2000STG} assume a non- asymptotically flat space-time  modeling the background of a galaxy, which could eventually explain high velocities of test particles in the center of a galaxy and at the same time the flat rotation curves. This sort of models assume there is a non-zero background energy density in the whole space-time which is a property inconsistent with asymptotic flatness.

The standard approach appropriate for the study of the common dark matter candidates of the wimp type, modeled by collision-less matter near supermassive black hole's region is focused on two major streams: i) explore the possibility that current supermassive black holes are the result of the growth through accretion by seed black holes of intermediate mass of about $10^3$ to $10^4 M_{\odot}$ of baryonic and dark matter \cite{seedBHs}, and ii) the contribution of dark matter to the growth of supermassive black holes. Concerning the second stream, some results indicate that only about 10 \% of their mass is due to the accretion of dark matter (e.g. \cite{freitas}), and more mature models involving the study of the Newtonian phase space \cite{losecone} indicate that the time-scale  for the accretion of collision-less dark matter to contribute significantly to a super massive black hole mass is too long (see e.g. \cite{Gilmore}). Also, important bounds on the collision-less dark matter central density have been proposed that allow the observed masses of central black holes without runaway instability of black holes, based on the study of orbits traveling on unbounded orbits with respect to the central object  \cite{HernandezLee}. More recent results show the importance of pressure of dark matter in the accretion rates, where it was shown that pressure itself is able to freeze the black hole's growth through dark matter accretion \cite{GuzmanLora2011}.

In this paper we proceed in a similar way, that is we study the evolution of a scalar field acting as dark matter around a supermassive black hole and study its behavior. We also make use of the approximation that considers that the dark matter is diluted and thus its evolution affects the black hole space-time in a minor way, thus allowing one to consider the dark matter to be a test matter field as assumed in those of the aforementioned studies. To our purpose it is particularly important noticing that a property of a test massive or massless scalar field evolving on a Schwarzschild space-time, is that it can only move toward two ends: the inside part of the black hole and future null infinity ({\it scri +}). That is, if the scalar field evolves around black holes, either supermassive black holes, or less massive black holes, we wonder how the scalar field behaves in time, whether it remains near the black hole, or is accreted by the black hole or it leaks toward {\it scri +}. 
Previous studies invoke the quantum nature of the scalar field to understand the accretion rate of SFDM into a black hole, that is, given that the scalar field is ultralight, it has a Compton wave-length much larger than the Schwarzschild radius and therefore the black hole will have a restricted cross section for such a scalar field (see for instance \cite{LiddleUrena}). Unlike such approach, in this paper we consider the scalar field to be classical as assumed in SFDM models at cosmic scales, from an effective Lagrangian:

\begin{equation}
{\cal L}=-R+ (\nabla \Phi)^2 + V(\Phi),
\end{equation}

\noindent where $R$ is the Ricci scalar of the space-time, $\Phi$ the scalar field and $V(\Phi)$ its potential. The variation of such Lagrangian with respect to $\Phi$ reduces to the Klein-Gordon (KG) equation, that rules the evolution of the scalar field, and since we work on a fixed background space-time there is no need to variate the Lagrangian with respect to the metric, which would imply Einstein's equations. 

Thus, we study numerically the behavior of a scalar field playing the role of SFDM on top of a black hole space-time. In order to study the evolution of the scalar field we assume the following conditions:
i) we set up a fixed Schwarzschild black hole space-time using hyperboloidal slices, and compactify the slices using coordinates such that our numerical domain contains {\it scri +},
 ii) the compactification of the spatial coordinate implies a singularity of the metric at future null infinity, thus we define a conformal space-time and solve the KG equation on top of such conformal space-time,
iii) we consider the scalar field to be a test field, that is, we do not solve for the evolution of the geometry of the space-time; this assumption works as a first approximation for a very diluted matter field,
iv) we solve the Klein-Gordon equation as a hyperbolic initial value problem, 
v) study the late time behavior of the scalar field and its independence on the initial data, so that we can determine an accretion-leak rate of the scalar field,
vi) we trigger the evolution of the scalar field by setting up initial data consisting on a time-symmetric shell of scalar field that presents both, an ingoing and an outgoing pulse,
vii) based on the study of the evolution in time of the scalar field amplitude we build up our conclusions.

Due to the considerable amount of calculations involved in the method we use, the units we assume in the paper are such that $G=c=1$ and the mass of the black hole $M=1$.

The paper is organized as follows, 
in section \ref{sec:2} we describe the construction of coordinates describing a Schwarzschild black hole using  constant mean curvature slices and scri-fixing compactification, in section \ref{sec:3} we describe the method used to solve the Klein-Gordon on the conformal space-time. In section \ref{sec:4} we present the evolution in time of the scalar field and interpret our results in terms of SFDM. Finally in section \ref{sec:5} we present some conclusions.


\section{Description of the compactified Schwarzschild space-time}
\label{sec:2}

An appropriate coordinate system of the Schwarzschild space-time to study the propagation of scalar fields can be achieved using hyperboloidal slices, because such slices reach future null infinity instead of spatial infinity, which is a natural boundary for a wave-like process, including electromagnetic fields and gravitational radiation. Additionally we require the slices to penetrate the event horizon of the black hole in order to track the accretion of the scalar field up to the interior of the black hole.

On the other hand, it is possible to compactify the space-time in order to work on a numerical domain that contains {\it scri +}, for which it is necessary to compactify the spatial coordinates in an appropriate way such that the space-time is regular there, at future null infinity. 
Usually, when a spatial coordinate is being compactified, the metric is singular at the new spatial infinity, and it also happens in the present case when hyperboloidal slices are compactified. Among others \cite{rinnemoncrief,rinne}, a  known way of fixing this problem is the definition of a conformal metric that uses an {\it ad hoc} conformal factor that regularizes the singular terms in the metric resulting from the compactification \cite{rinne,AlexLoraGuzman}. The idea behind  scri-fixing conformal compactification for spherically symmetric space-times is the following: i) use hyperboloidal foliations whose slices reach {\it scri +}, ii) compactify the radial coordinate, iii) regularize the metric with an appropriate conformal rescaling. 

Assuming a spherically symmetric space-time is labeled using coordinates $(\tilde t, \tilde r,\theta,\varphi)$, a well known process to foliate the space-time using hyperboloidal slices starts by defining a new time coordinate $t = \tilde{t} - h(\tilde{r})$, where $h(\tilde{r})$ is a function called height function. This transformation has the advantage that keeps the time direction invariant. 
A compactifying coordinate is introduced in the form  $\tilde{r} =  \frac{r}{\Omega}$, where $\Omega=\Omega(r)$ is a function that compactifies $r$ and at the same time is a non-negative conformal factor that vanishes at  {\it scri +}; in the present case we choose $\Omega=1-r$ and then {\it scri + } is  located at $r=1$. We rescale the space-time using a conformal transformation $g=\Omega^{2}\tilde{g}$ so that the conformal metric $g$ is regular everywhere because the conformal rescaling compensates the singularities in the metric due to the compactification of the radial coordinate \cite{AlexLoraGuzman}.

For the construction of $h(\tilde{r})$, constant mean curvature slices (CMC) are used \cite{Murc}. The mean extrinsic curvature $\tilde{k}$  of the initial slice $t=0$ is given by 
$\tilde{k} = \nabla_{\mu}n^{\mu} = \frac{1}{\sqrt{-g}}\partial_{\mu}(\sqrt{-g}n^{\mu})$, 
where $n^{\mu}$ is a time-like unit vector normal to the spatial hypersurface and positive $\tilde{k}$ means expansion. In this case, we adopt the requirement that the mean extrinsic curvature $\tilde{k}$ is constant. This condition allows the integration of the equation above. For the Schwarzschild
metric
 
\begin{equation}
d\tilde{s}^2 = -\alpha^2(\tilde{r}) d\tilde{t}^2 + a^2(\tilde{r}) d\tilde{r}^2 +
	\tilde{r}^2 (d\theta^2 + \sin^2 \theta d\varphi^2),
\label{eq:Boson-metric}
\end{equation}

\noindent where $\alpha = 1/a = \sqrt{1-\frac{2}{\tilde r}}$ it is found that, after solving for $h'$ the condition $\tilde k=constant$,  the final version of the conformal metric is \cite{Murc}:

\begin{equation}
ds^{2}=-\left(1-\frac{2\Omega}{r}\right)\Omega^{2}dt^{2}-\frac{2(\tilde{k}r^{3}/3 -C\Omega^{3})}{P(r)}dtdr 
+ \frac{r^{4}}{P^{2}(r)} dr^{2}+ r^{2}(d\theta^{2} + \sin^2 \theta d\varphi^{2}),
\label{eq:Sch_compactified}
\end{equation}

\noindent where

\begin{equation}\label{eq:ppp}
P(r)=\Omega^{3}\tilde{P}(r)= \sqrt{\left(\frac{\tilde{k}r^{3}}{3}-C\Omega^{3}\right)^{2} + \left(1-\frac{2\Omega}{r}\right)\Omega^{2}r^{4}}.
\end{equation}

\noindent The values of $\tilde{k}$ and the integration constant $C$ are restricted such that $P(r)$ is real; the units of $C$ are those of $M^2$. The values of these constants  used to obtain our are $\tilde{k} = 0.4$ and $C=2$ in all our calculations, and it has been shown that for the massless case the results are independent of the values of these constants \cite{AlexLoraGuzman}.


\section{Klein-Gordon equation: solution and numerical aspects}
\label{sec:3}

The Klein-Gordon equation for a scalar field $\tilde\phi_T$ reads:
$\tilde{\Box} \tilde \phi_T - \frac{d \tilde V}{d \tilde \phi_T} = 0$, where $\tilde \Box
\tilde \phi_T=\frac{1}{\sqrt{- \tilde g}}\partial_{\mu}[\sqrt{- \tilde g}
\tilde g^{\mu\nu}\partial_{\nu} \tilde \phi_T]$, and we consider  the potential to be of the form $\tilde V=\frac{1}{2}m_{B}^{2} |\tilde \phi_T|^2 + \frac{\lambda}{4}|\tilde \phi_T|^4$, where $m_B$ has units of mass. The KG equation in the conformal metric reads

\begin{equation}
\tilde{\Box} \tilde{\phi}_T - \frac{1}{6} \tilde{R}\tilde{\phi}_T -(m_B^{2}\tilde{\phi}_T + \lambda \tilde{\phi}_T^3) =
\Omega^{3}\left[\Box \phi_T - \frac{1}{6} R\phi_T -(m_B^{2} \Omega^{-2}\phi_T + \lambda \phi_T^3)\right] =0,\label{eq:conf}
\end{equation}

\noindent provided the relationship between the physical scalar field $\tilde \phi_T$ and the conformal scalar field $\phi_T$ is $\phi_T=\tilde \phi_T/ \Omega$. Here $R=\frac{12\Omega}{r^2}(r + (2r-1))$ is the Ricci scalar of the conformal metric and $\Box=\nabla^{\mu}\nabla_{\mu}$ corresponds to the conformal metric. The case  for which equation (\ref{eq:conf}) is conformally invariant corresponds to the zero mass case $m_B =0$, however, as we want to study the behavior of the test scalar field on a fixed Schwarzschild background, we solve the conformal KG equation that contains the mass parameter different from zero and avoid the future null infinity boundary at $r=1$ in our calculations and we only approach such boundary in order to consider a big spatial physical domain. 

The scalar field is separated in the form $\phi_T (t,r,\theta,\varphi) = \phi(t,r)Y_{lm}(\theta,\varphi)$ for a given value of $l$, in order to explore the behavior of angular components of the scalar field in the process.  As mentioned above, since the conformal KG equation (\ref{eq:conf}) is singular at $r=1$, we do not have  the chance to solve the equation in the whole spatial domain and we are forced to set up an artificial boundary  close to {\it scri +}  approximately at a distance $1000$ in units of the mass of the black hole. We also implement an expansion of the $\Omega^{-2}$ factor in (\ref{eq:conf}) in a power series $\sum_{n=0}^{\infty} r^n$. At such boundary we apply outgoing wave boundary conditions, and since the boundary is pretty close to {\it scri +} the ingoing velocity is nearly zero and no numerical incoming signals affecting our calculations were observed. 

We implement another boundary inside the black hole's horizon at $r_{min}$, a method called excision. Notice first that the black hole's horizon according to (\ref{eq:Sch_compactified}) is located at $r=2/3$. For our calculations we impose the boundary at $r_{min}=0.6$, where we know the light cones point toward the singularity and no signals are propagated outward, which also involves spurious signals due to numerical errors, this is why there is no need to impose boundary conditions at this excision boundary.

We solve (\ref{eq:conf}) as an initial value problem using a first order formulation and usual ADM gauge functions. In terms of the line element (\ref{eq:Sch_compactified}) we can read off the gauge in terms of the ADM-like metric $d\hat{s}^2 = (-\hat{\alpha}^2 + \hat{\gamma}^2 \hat{\beta}^2)dt^2 + 2 \hat{\beta}\hat{\gamma}^2 drdt + \hat{\gamma}^2 dr^2 + r^2 (d\theta^2 + \sin^2 \theta d\varphi^2)$ and obtain the following gauge and metric functions

\begin{eqnarray}
&&\hat{\alpha}^{2}(r) = \alpha^2(r) \Omega^2 + \hat{\beta}^2(r) \hat{\gamma}^2(r),\nonumber\\
&&\hat{\beta}(r) = -\frac{\alpha^2(r) h'(r) (\Omega - r\Omega')}{\hat{\gamma}^2(r)},\nonumber\\
&&\hat{\gamma}^2(r) = \frac{(a^2(r) - \alpha^2(r) h'^2(r))(\Omega-r\Omega')^2}{\Omega^2},
\label{eq:gauge_sch}
\end{eqnarray}

\noindent which are the final metric functions we use to solve the KG equation. For numerical convenience we solve the KG equation as a first order system, for which we  define first order variables $\pi = \frac{\hat \gamma}{\hat \alpha}\partial_t \phi_T - \frac{\hat \gamma}{\hat \alpha} \hat \beta \partial_r \phi_T$ and $\psi = \partial_r \phi_T$. In terms of these new variables the KG equations splits into the following  system of first order equations (see the appendix for details)

\begin{eqnarray}
&&\partial_t \psi=  \partial_r\left( \frac{\hat \alpha}{\hat \gamma}\pi + \hat \beta \psi \right),\nonumber\\
&&\partial_t \pi = \frac{1}{r^2} \partial_r \left(r^2 (\hat \beta \pi + \frac{\hat \alpha}{\hat \gamma}\psi ) \right) - 
\hat \alpha \hat \gamma \left( \frac{1}{6}R\phi_T + \frac{l(l+1)}{r^2}\phi_T + m_B^{2} \Omega^{-2}\phi_T + \lambda \phi_{T}^{3} \right),\nonumber\\
&&\partial_t \phi_T = \frac{\hat \alpha}{\hat \gamma} \pi + \hat \beta \psi, \nonumber \\
&&\partial_r \phi_T = \psi,
\label{eq:1st_order}
\end{eqnarray}

\noindent where the first two equations correspond to the evolution of the first order variables, the third one is the definition of $\pi$ that we use to reconstruct the value of $\phi_T$, and the last equation is a constraint that has to be satisfied during the evolution. We have included the contribution of the angular part of the scalar field. System (\ref{eq:1st_order}) is a symmetric hyperbolic system, which implies that we start up with a well posed initial value problem \cite{Thomas}.

We integrate system (\ref{eq:1st_order}) numerically using a finite differences approximation on a uniformly discretized domain, we use the method of lines with sixth order accurate stencils along the spatial direction and a fourth order Runge-Kutta time integrator. In all our runs we use a constant Courant-Friedrichs-Lewy factor (CFL), such that $\Delta t = CFL \times \Delta x$, where $\Delta t$ and $\Delta x$ are respectively the time and spatial resolutions of our discretized domain. In order to set initial data consistent with (\ref{eq:1st_order}) we restrict to a particular set of cases. That is, for the case $\lambda=0$ eq. (\ref{eq:1st_order}) is the same as for $\phi(t,r)$ and $l=0,1,2,...$, because the spherical harmonics are factored out, for this reason initial data for $\phi(t,r)$ and any value of $l$ is consistent. On the other hand, for the case $\lambda \ne 0$, eq. (\ref{eq:1st_order}) depends on the spherical harmonics due to the $\phi_{T}^{3}$ term; in order to simplify our parameter space we choose to provide only initial data for $\phi(t,r)$ and only the case $l=0$ for which spherical harmonics are trivial and the initial data are consistent. 
We use initial data corresponding to a time-symmetric Gaussian profile for the scalar field because it includes both, a mode moving toward the hole and an outgoing pulse of scalar field. The initial data used then read

\begin{eqnarray}
\phi(0,r) &=&  Ae^{-(r-r_{0})^2/\sigma^2}, \nonumber \\
\psi(0,r) &=& -2 \frac{(r-r_{0})}{\sigma^2}\phi(0,r),\nonumber\\
\pi (0,r) &=& 0, \label{eq:initial_data}
\end{eqnarray}

\noindent and explore the parameter space for the amplitude and width of the Gaussian profile, while we keep $r_0$ fixed at $r_0=0.8$ which in the physical radial coordinate means $\tilde r_0 = 4$, and is well out of the event horizon. The value of $\sigma$ is extended to cover various length scales, given that for $r>0.8$ the physical coordinate $\tilde r$ grows rapidly as $\tilde r = \frac{r}{1-r}$.


\section{Results}
\label{sec:4}

In this part, unless otherwise stated we use initial scalar field profiles with $m_{B}^{2}=0.0, 0.1, 0.2$. 
In order to explore the parameter space, we use various values of the amplitude $A=0.01,0.1,1.0$ and different widths of the Gaussian pulse, which in physical units correspond to $\sigma_1=0.5 $, $\sigma_2= 1 $ and $\sigma_3 = 5 $ (in units of the mass of the black hole) to the right from $r_0=0.8$. This range of parameters allows us to cover length ranges that involve potential Compton wave-length related effects of interaction and reflection suggested recently \cite{mendozza}, that may involve reflection and absorption effects.

\begin{center}
\begin{figure}[htp]
\includegraphics[width=6cm]{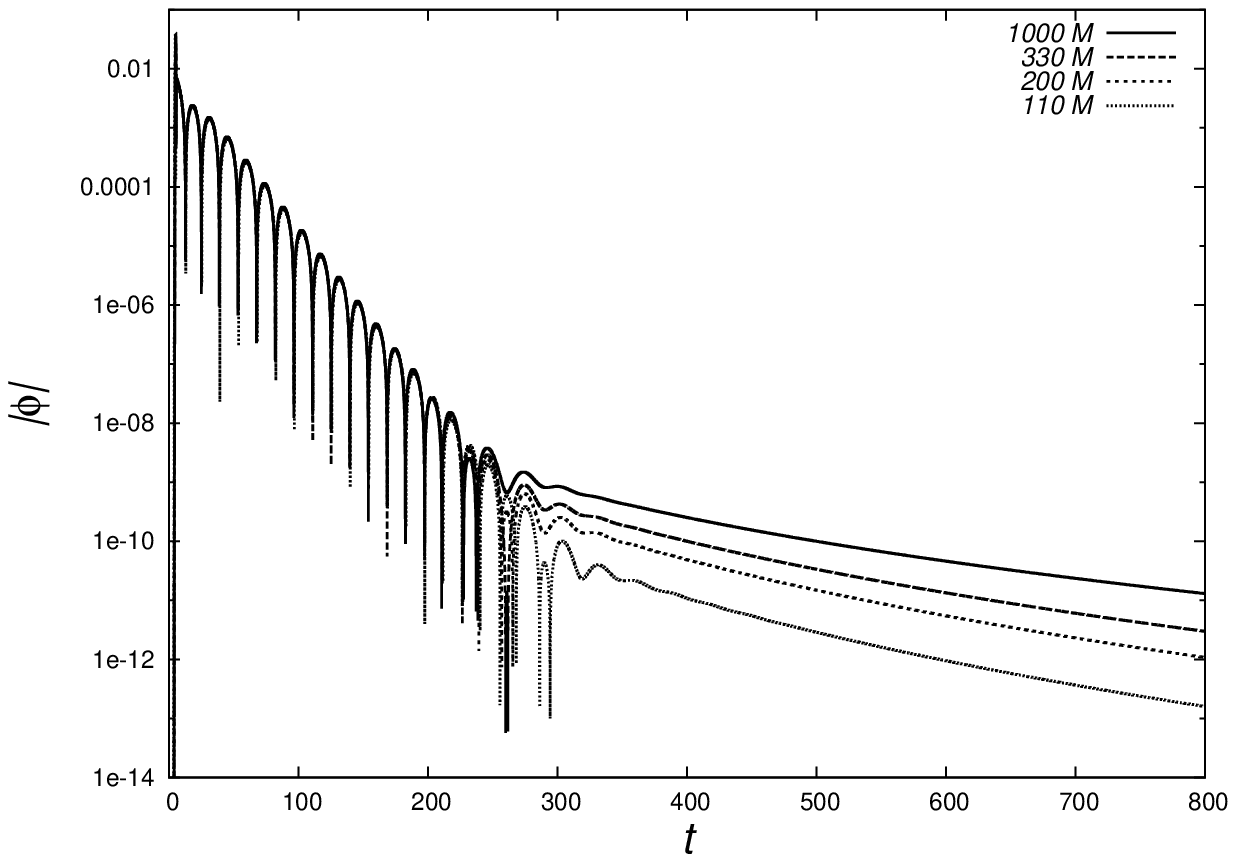}
\includegraphics[width=6cm]{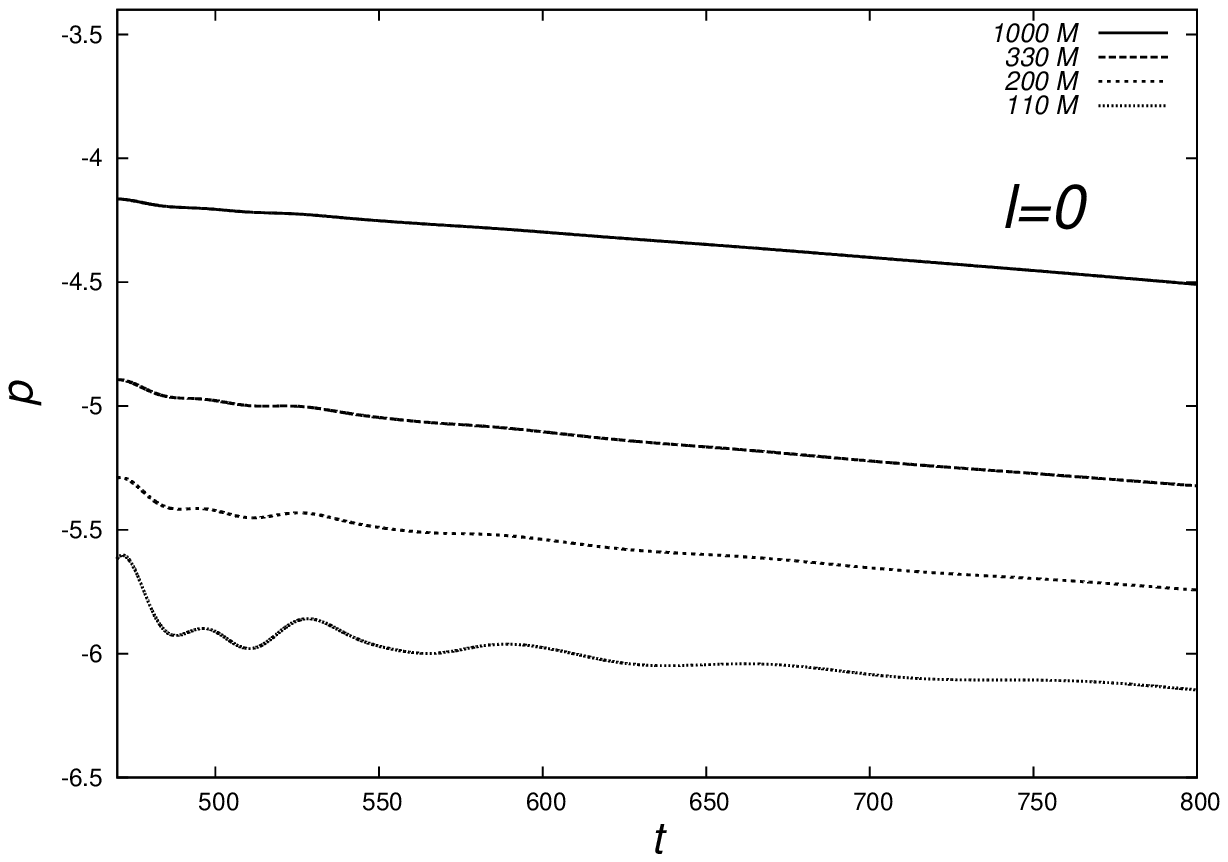}
\includegraphics[width=6cm]{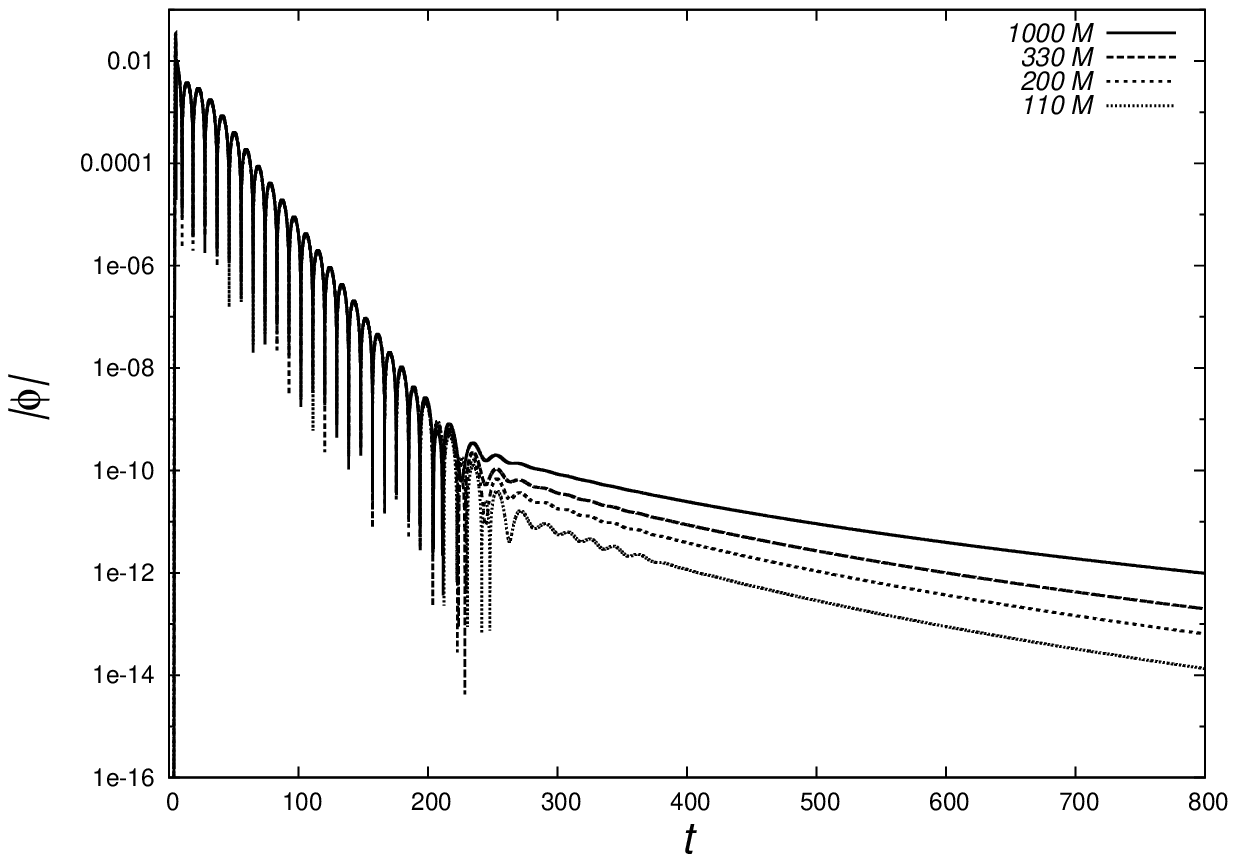}
\includegraphics[width=6cm]{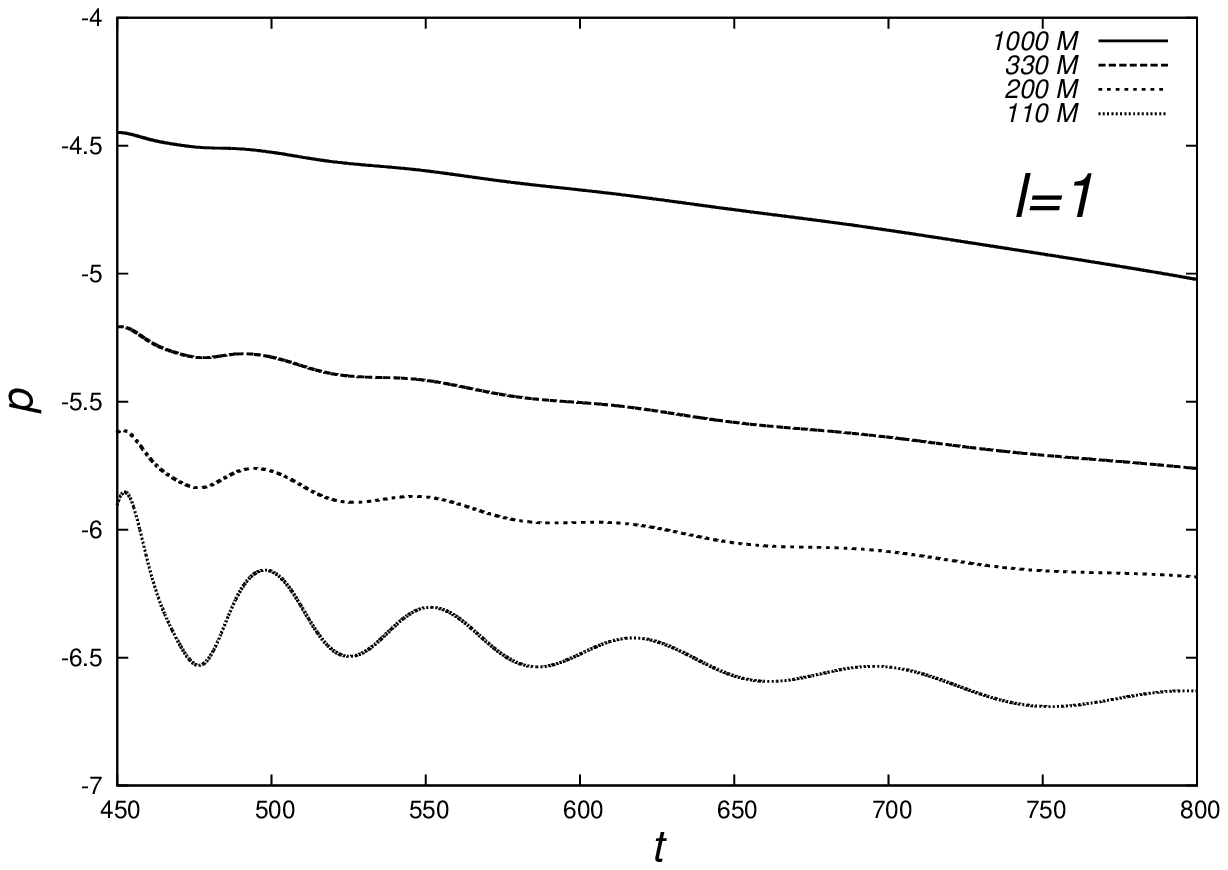}
\includegraphics[width=6cm]{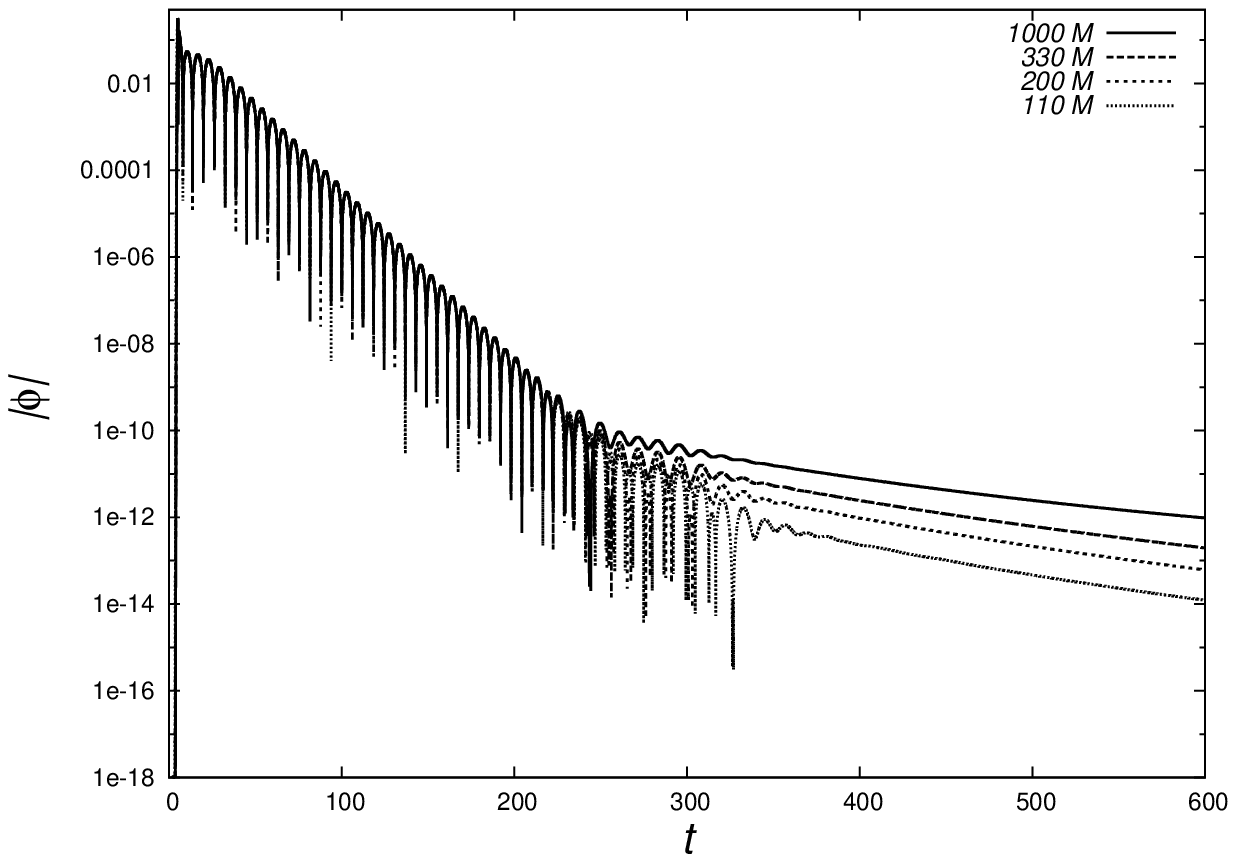}
\hspace{3.4cm} \includegraphics[width=6cm]{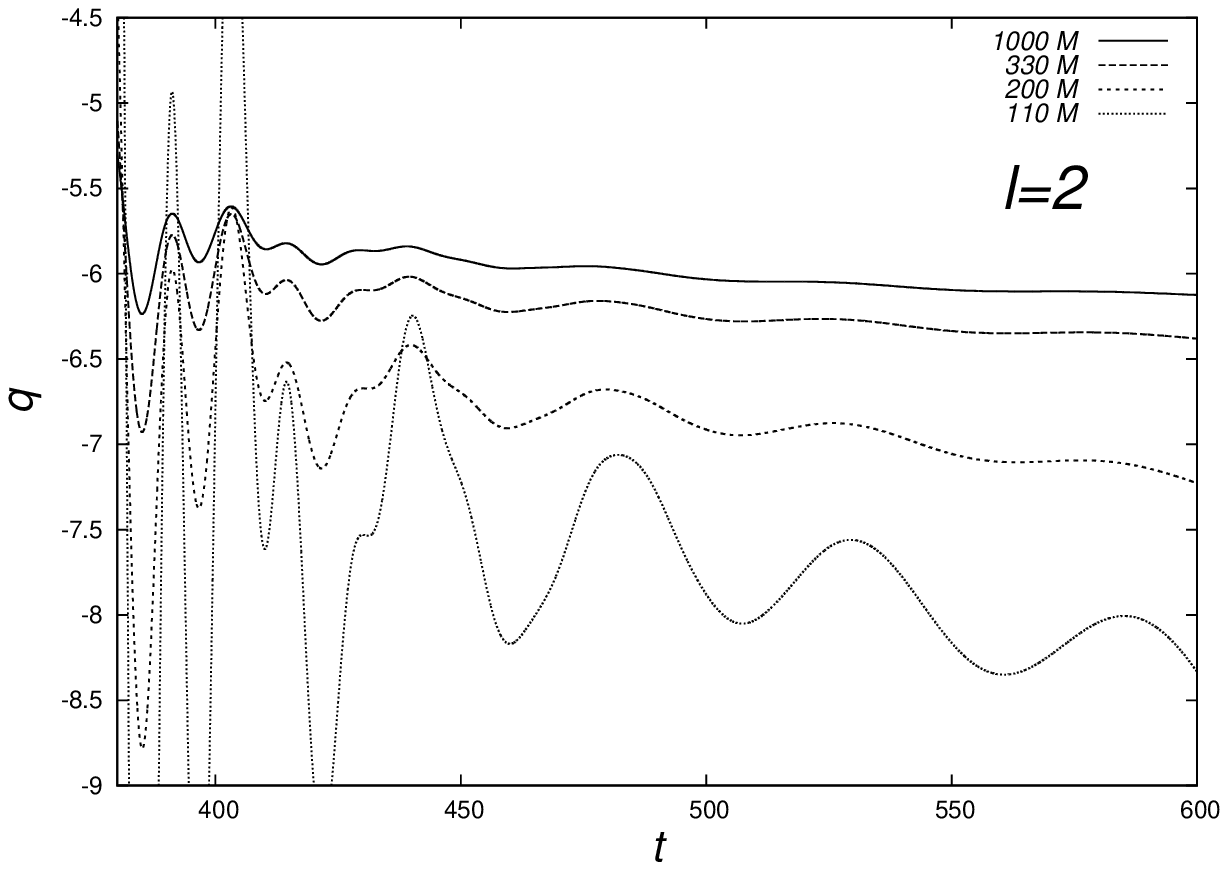} 
\caption{\label{fig:detectors} (Left) We show the amplitude of the massive scalar field in time measured at various distances from the black hole's horizon for $l=0,1,2$ and $m_B^{2} = 0.1$. The time dependence shows an oscillatory stage similar to that found for the massless case followed by a polynomial-like decay rate $\sim t^p$ with $p<0$; each line indicates the scalar field amplitude measured at a different distance from the black hole in units of the mass of the black hole. (Right) We show the exponent $p$ for the various cases. We show that the decay rate of the scalar field is bigger with bigger values of $l$, that is, the amplitude decreases faster during the tail-like stage when non-trivial angular components of the scalar field get involved. If this scalar field is to be considered to be the dark matter, this indicates that the amplitude decays faster if the scalar field has angular components.}
\end{figure}
\end{center}

\begin{figure}[htp]
\includegraphics[width=6cm]{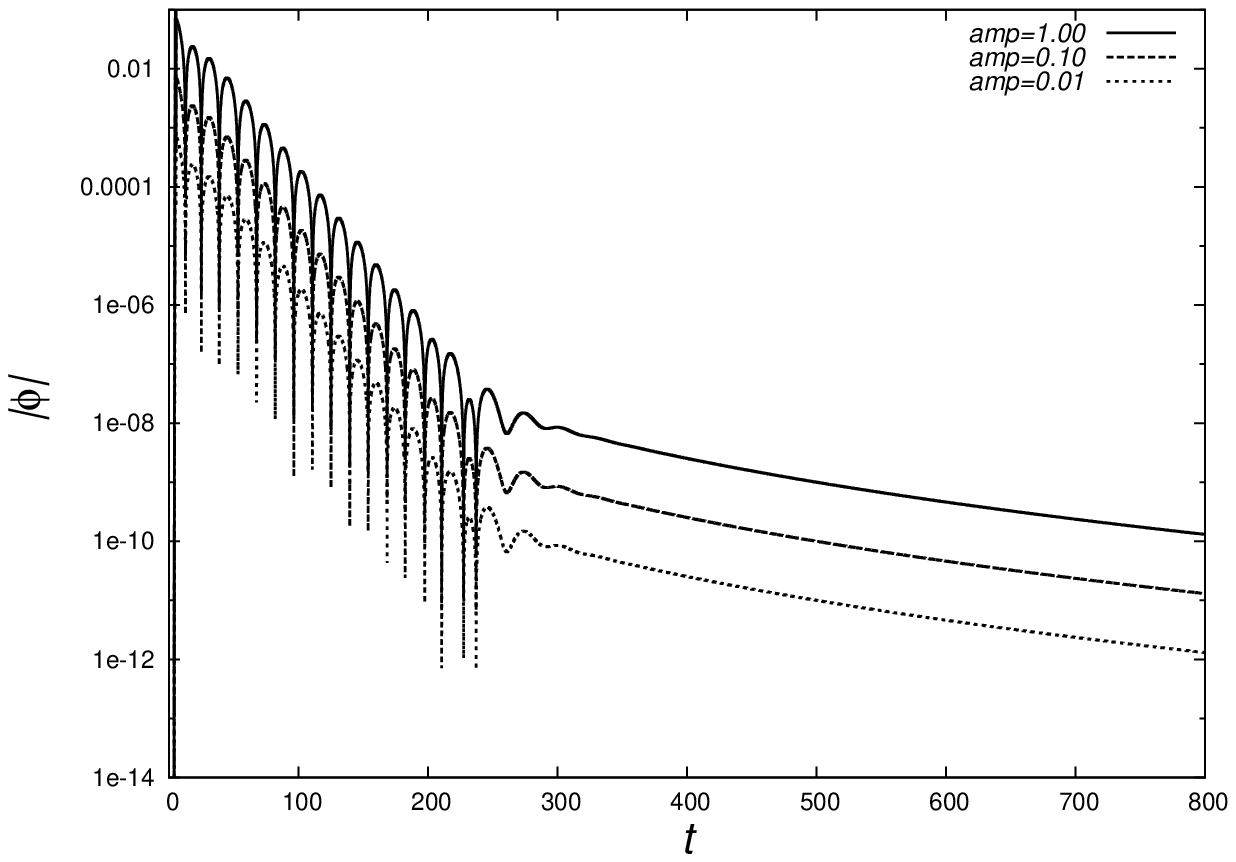}
\includegraphics[width=6cm]{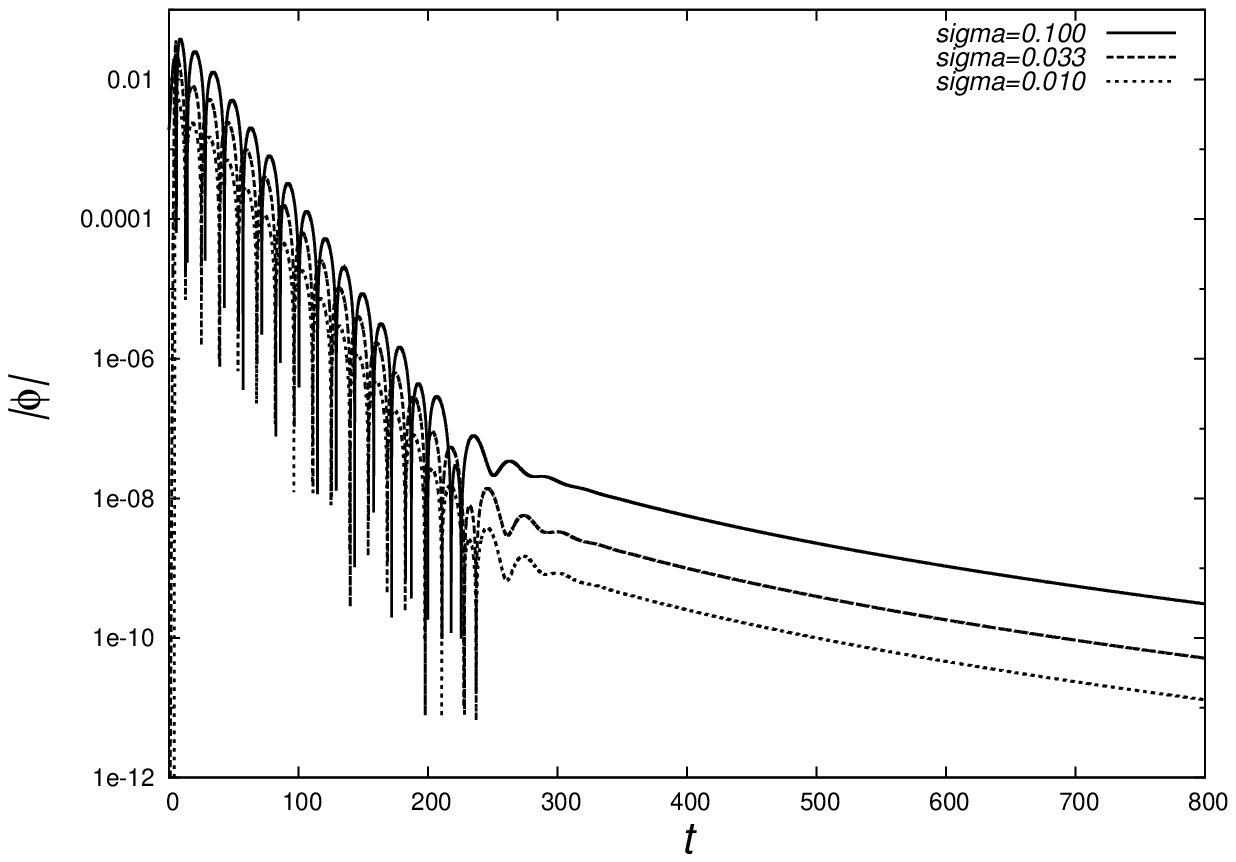}
\includegraphics[width=6cm]{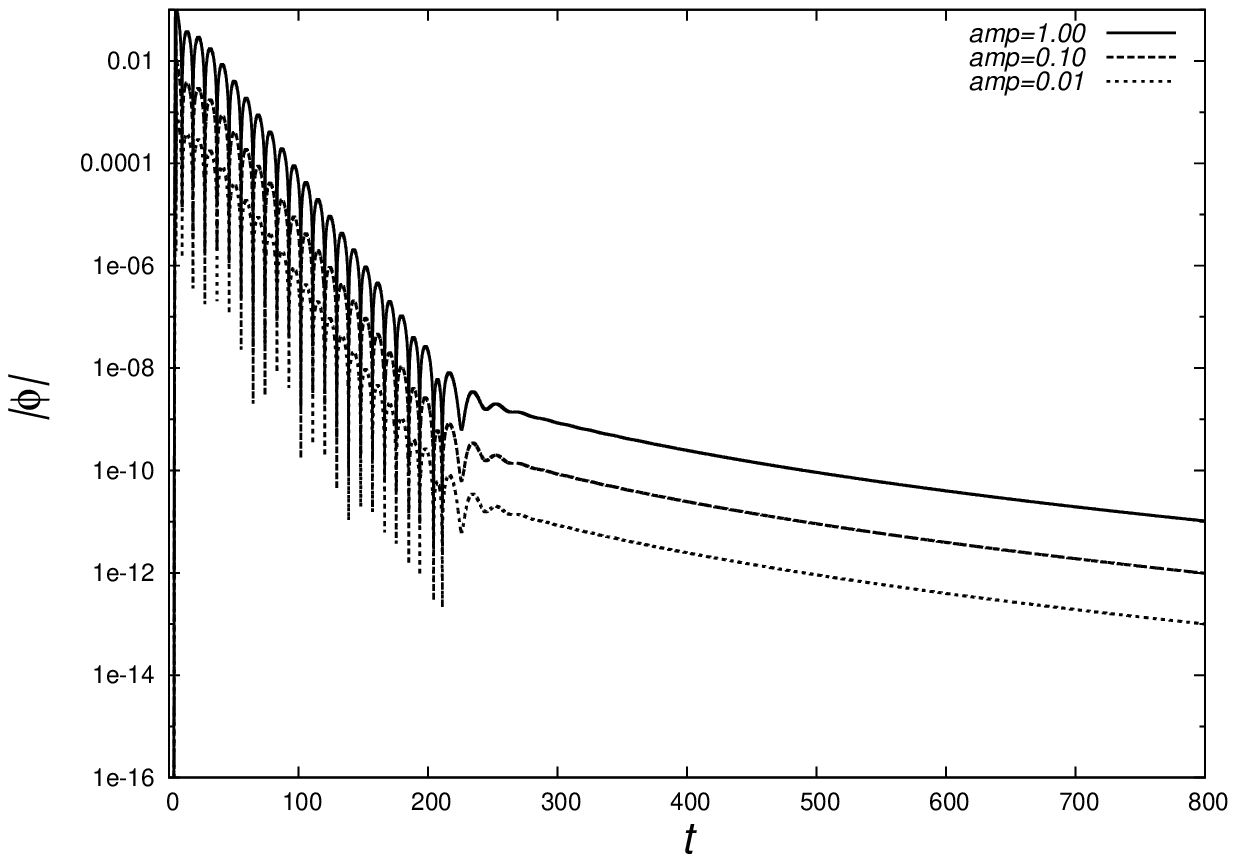}
\includegraphics[width=6cm]{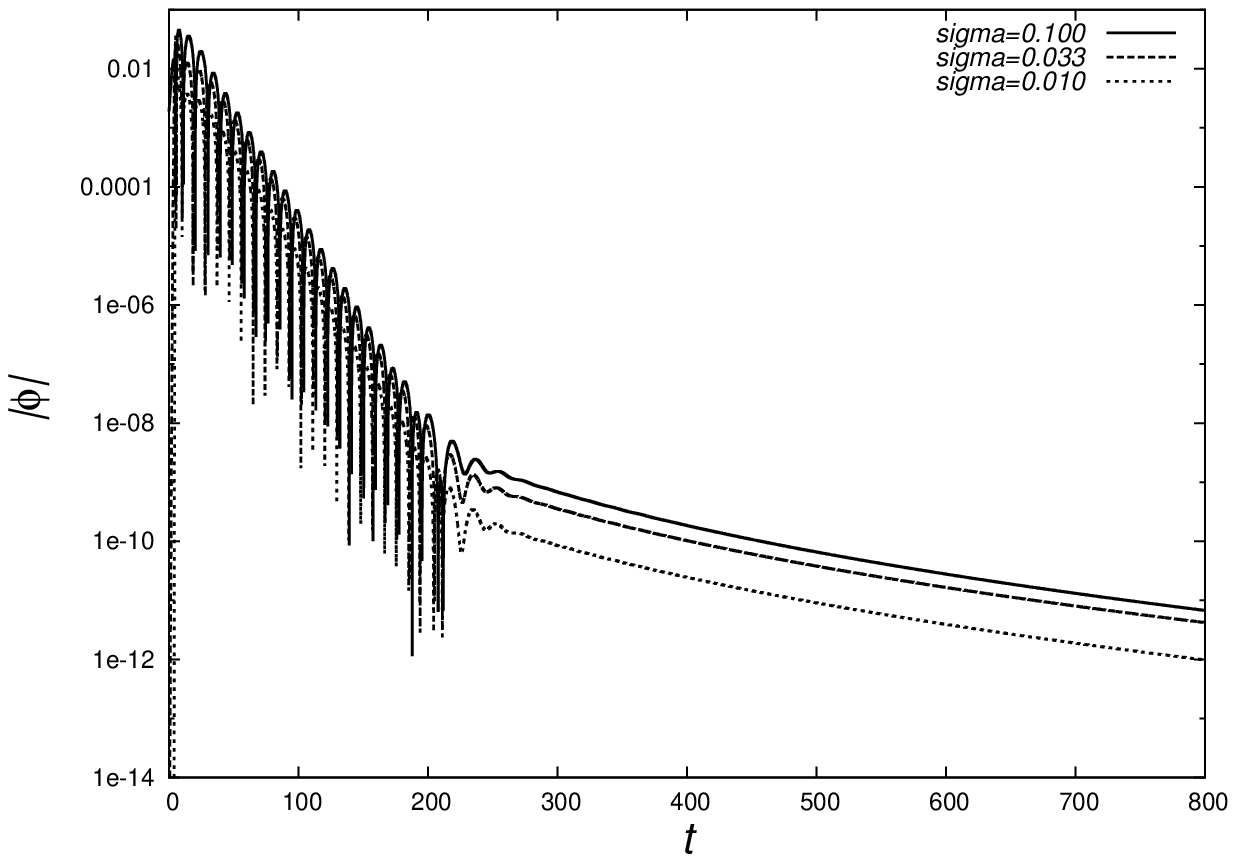}
\includegraphics[width=6cm]{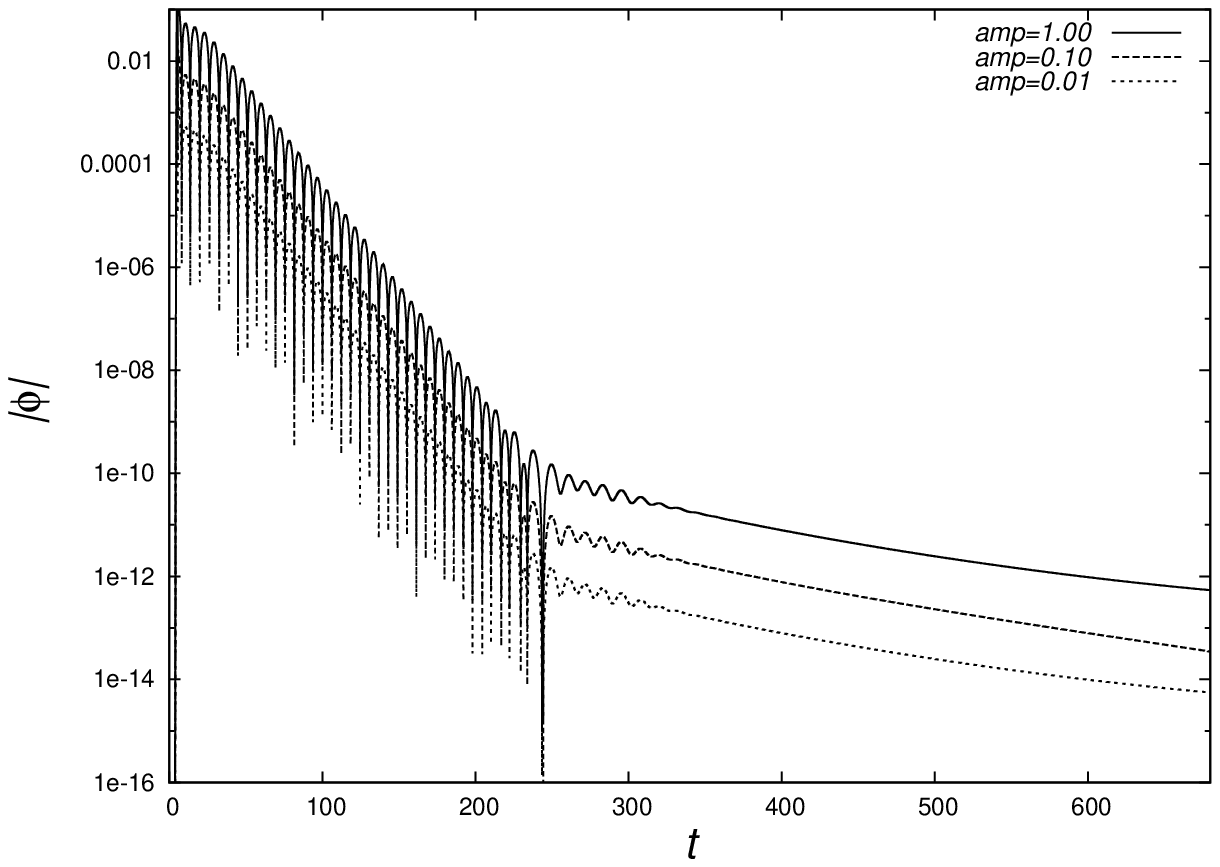}
\hspace{3.4cm} \includegraphics[width=6cm]{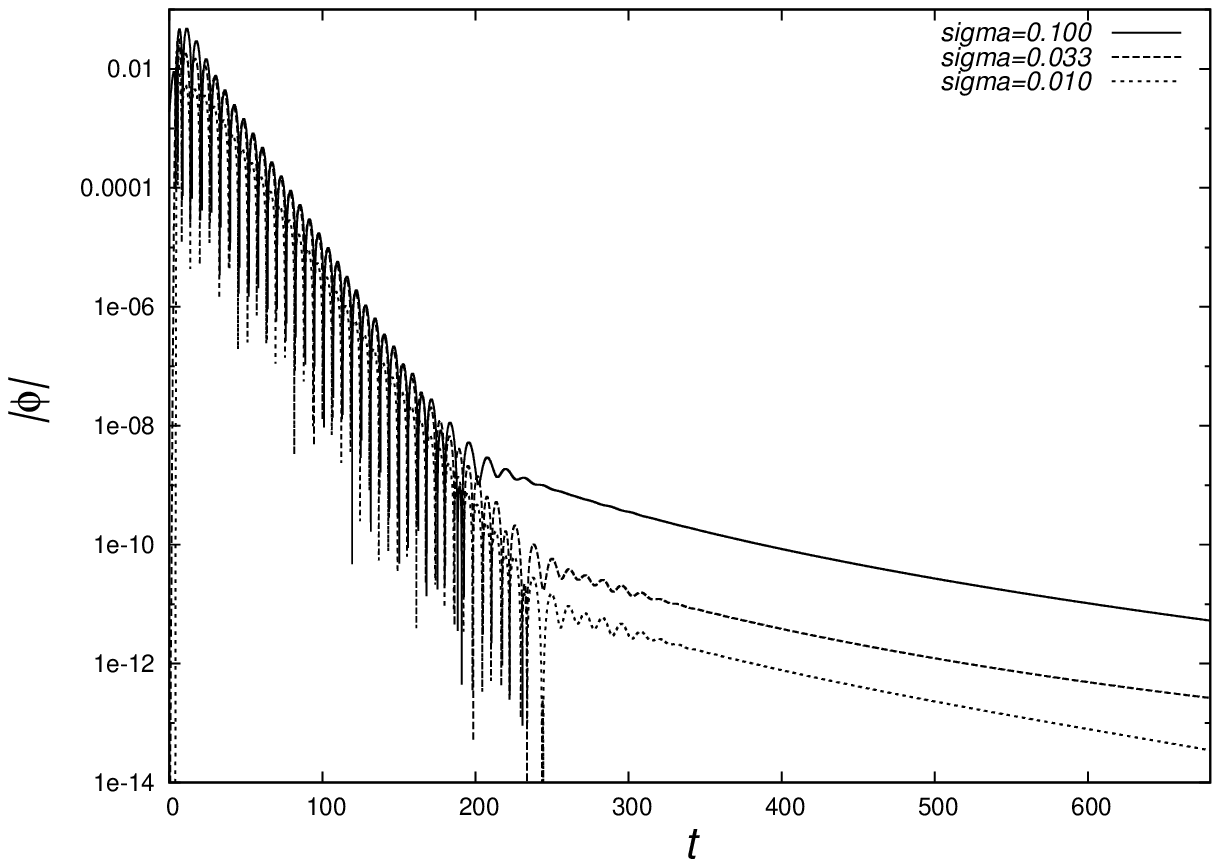}
\caption{\label{fig:parameters} In these figures, we show the amplitude of the massive scalar field measured at $1000$ from the black hole's horizon, and  for different values of the amplitude $A$ and widths $\sigma$ of the Gaussian. We show the evolution for three values of the angular term $l=0,1,2$ (from top to bottom respectively) and scalar mass $m_B^{2}=0.1$. We found that the tail behavior of the amplitude is about the same. This implies that scalar field decays at same rate independently of the profile  parameters.}
\end{figure}

The evolution is tracked during a time window in which our results converge and that also allow to track the amplitude of the scalar field from the quasinormal mode ringing stage (in the massless scalar field case these are actually the quasinormal mode oscillations of the black hole) where the amplitude measured at various distances from the horizon behaves as $|\phi| \sim e^{i(\omega_R + i\omega_I)t}$, up to a polynomial tail decay stage of the amplitude, during which the time dependence of the amplitude is of the type $|\phi| \sim t^p$ with $p<0$.

We measure the amplitude of the scalar field at various distances from the black hole's horizon. In Fig. \ref{fig:detectors} we show the evolution of its amplitude, and find as expected an oscillatory lapse followed by an approximately polynomial tail decay (the exponent does not stabilize around a constant value) for various locations of the detectors and three different values of $l$. We also show the exponent of the polynomial decay $p$ in each case. We notice that the amplitude decays with decreasing exponents between $-6.5<p<-4$, depending on the location of the detector for the $l=0$ case. For the cases with non trivial angular component the decay is even faster, for instance for the $l=2$ case the exponent lies between $-9<p<-5$ even if such exponent has not yet stabilized. In these results we only show the time window within which our calculations converge at least with fourth order, which is the accuracy order of the time integrator as shown below.

In Fig. \ref{fig:parameters} we show the time dependence of the scalar field amplitude for various values of the parameters of the initial data. The finding is that the tail behavior is independent of the width and amplitude of the initial gaussian we explored, which indicates a universal behavior which does not seem to depend on the initial data. That at the tail stage of the amplitude the exponent of the decay is independent of the amplitude and width of the Gaussian profile allows us to extract conclusions for an initial SFDM packet represented by the scalar field.

\begin{center}
\begin{figure}[htp]
\includegraphics[width=8cm]{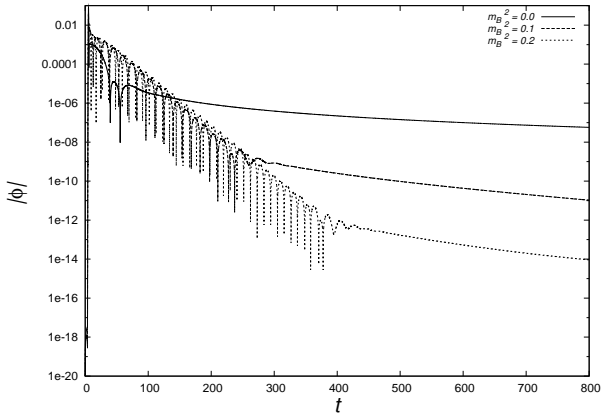}
\caption{\label{fig:mass} We present the evolution of the scalar field amplitude for different values of the scalar field mass $m_{B}^{2}$ measured by a detector located at $1000$ in units of the mass of the black hole, for the $l=0$ case. We found that  the decay rate increases with the mass of the scalar field.}
\end{figure}
\end{center}

In Fig.\ref{fig:mass} we show the amplitude of the scalar field for different values of the boson mass $m_{B}^{2}$, measured by a detector located at $r=1000$ for $l=0$. We were unable to explore the parameter space for bigger values of $m_{B}^{2}$ because the tail-like decay appears at considerably small amplitudes near the round-off error threshold of our calculations, and we were unable to achieve convergence at such regimes, however this figure suggests a trend: the bigger the values of $m_{B}^{2}$ the faster the amplitude of the scalar field decays; this is consistent with the fact that the massive term in our evolution equations is always switched on and acts as a dissipative term.

\begin{figure}[htp]
 \includegraphics[width=6cm]{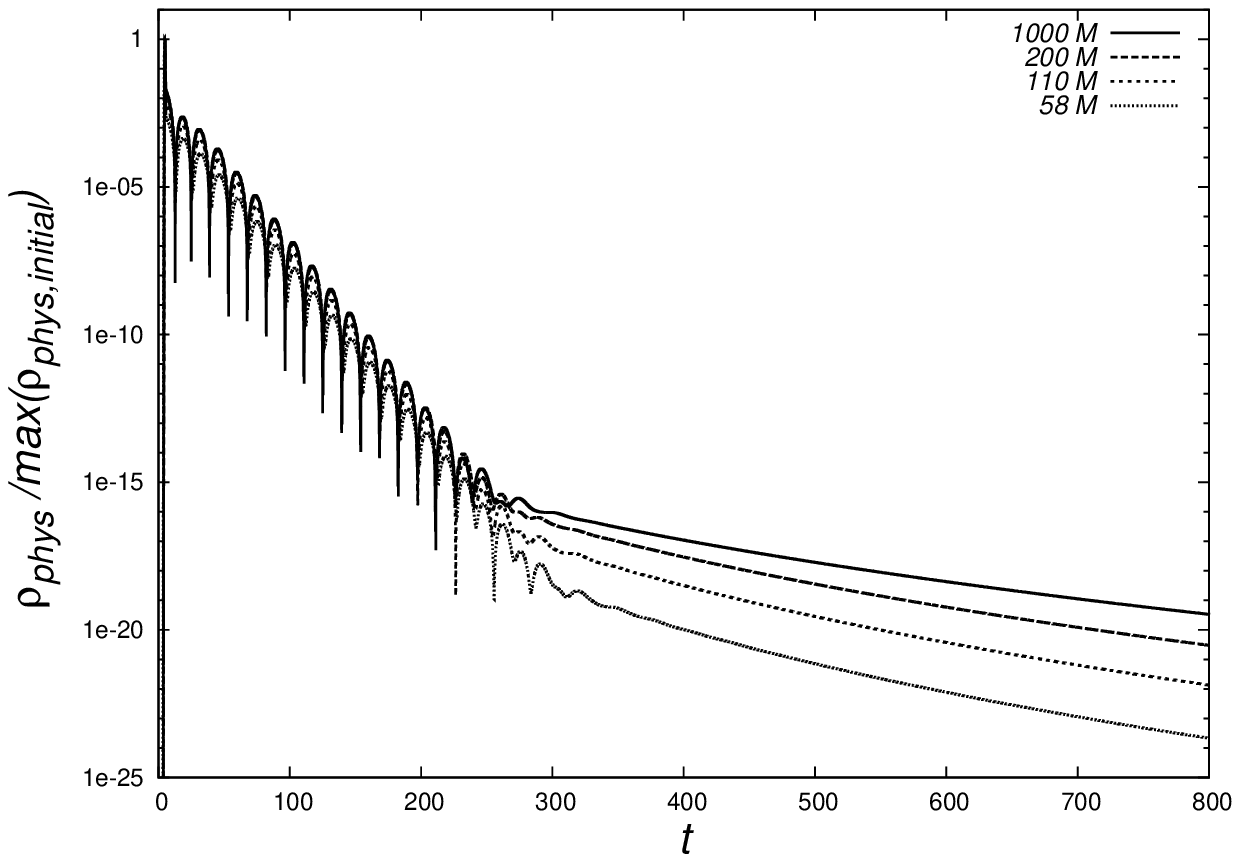}
 \includegraphics[width=6cm]{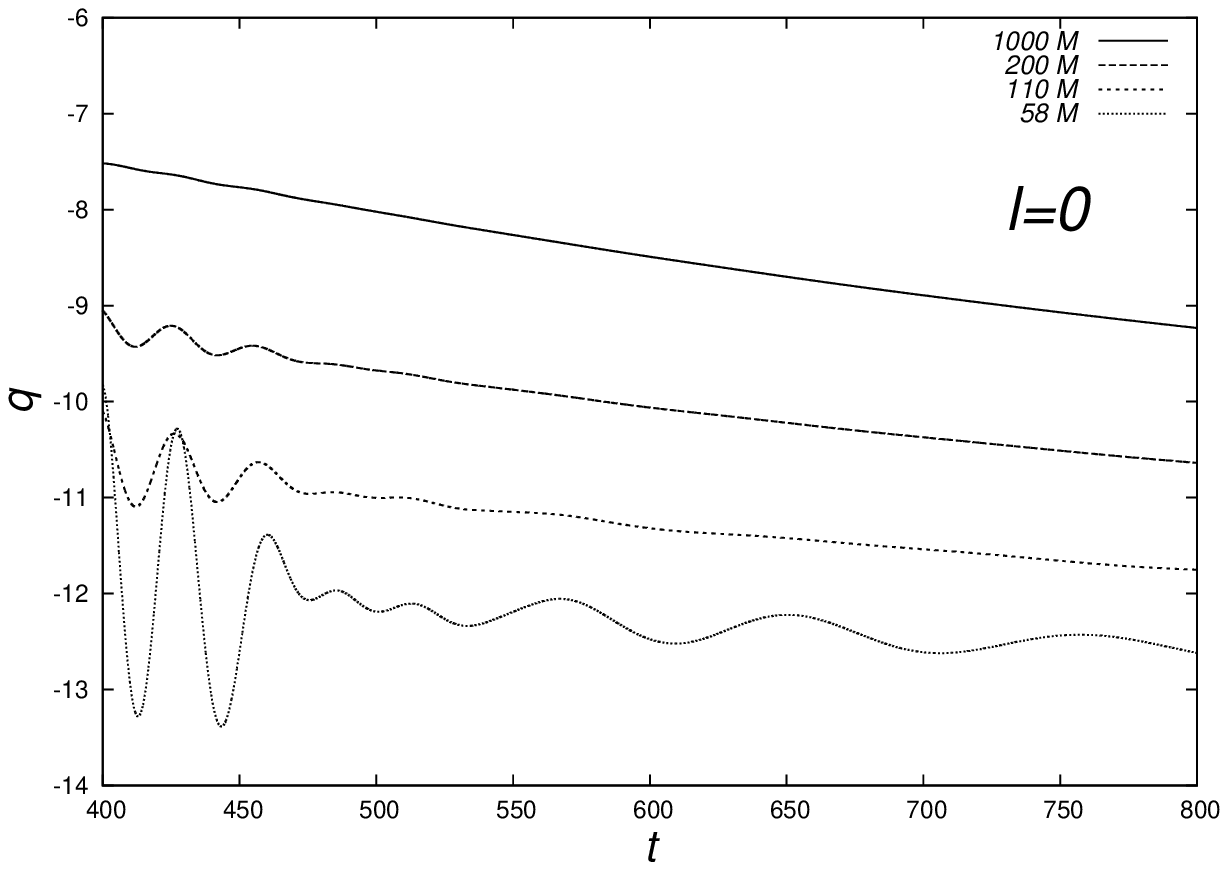}
 \includegraphics[width=6cm]{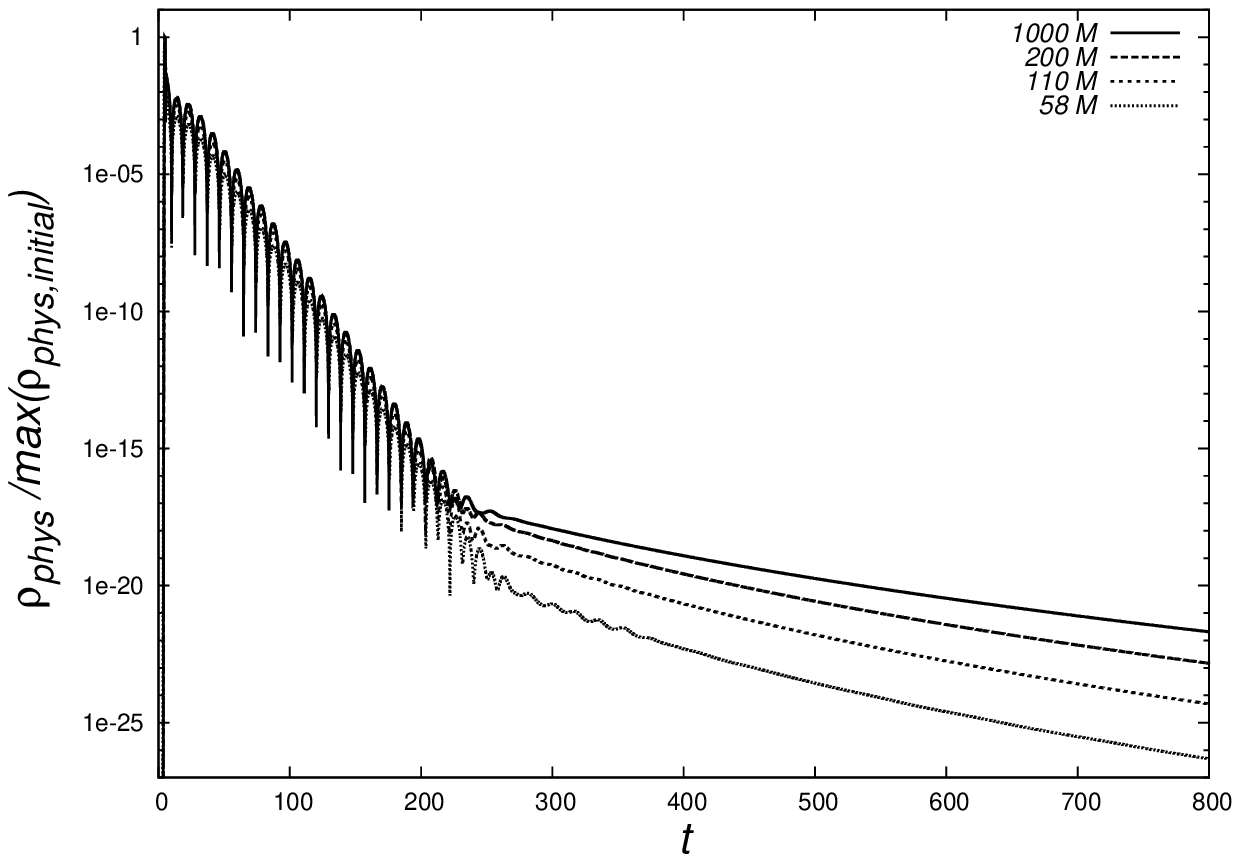}
 \includegraphics[width=6cm]{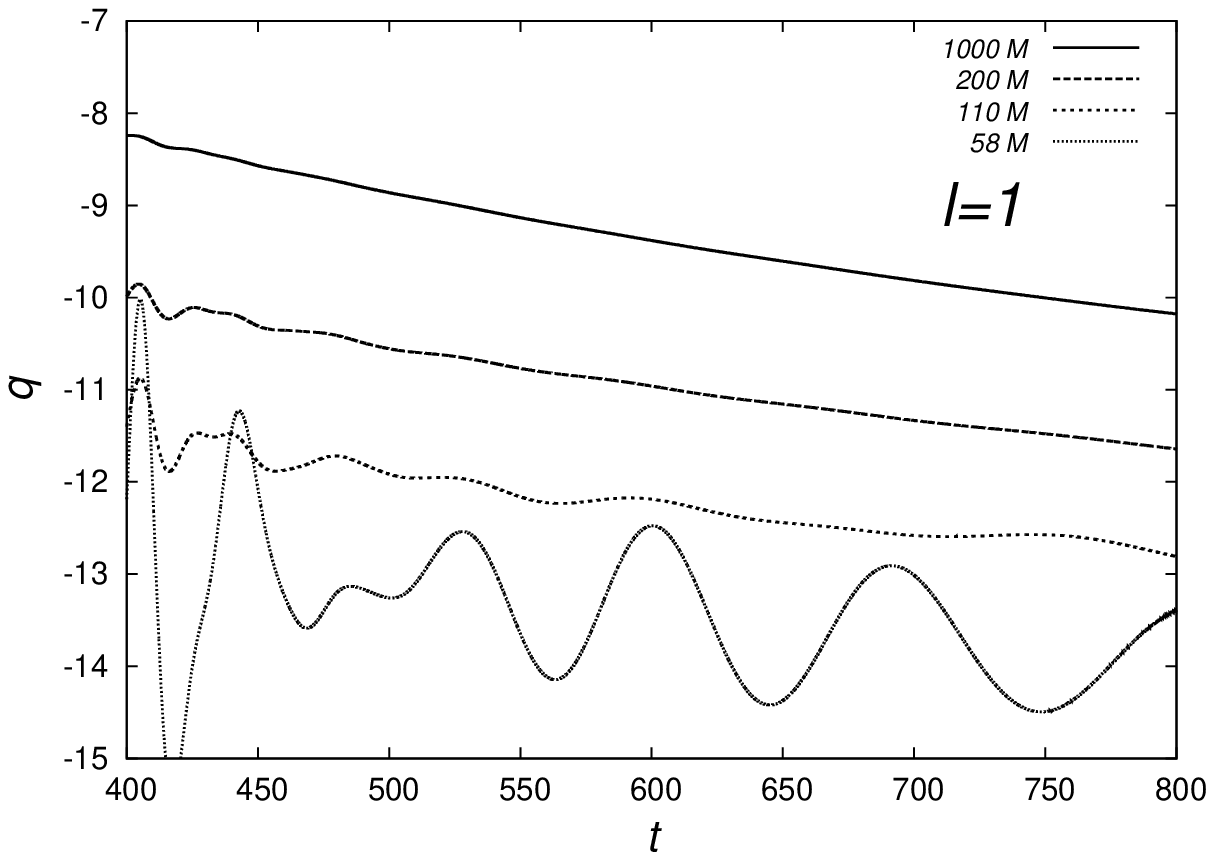}
 \includegraphics[width=6cm]{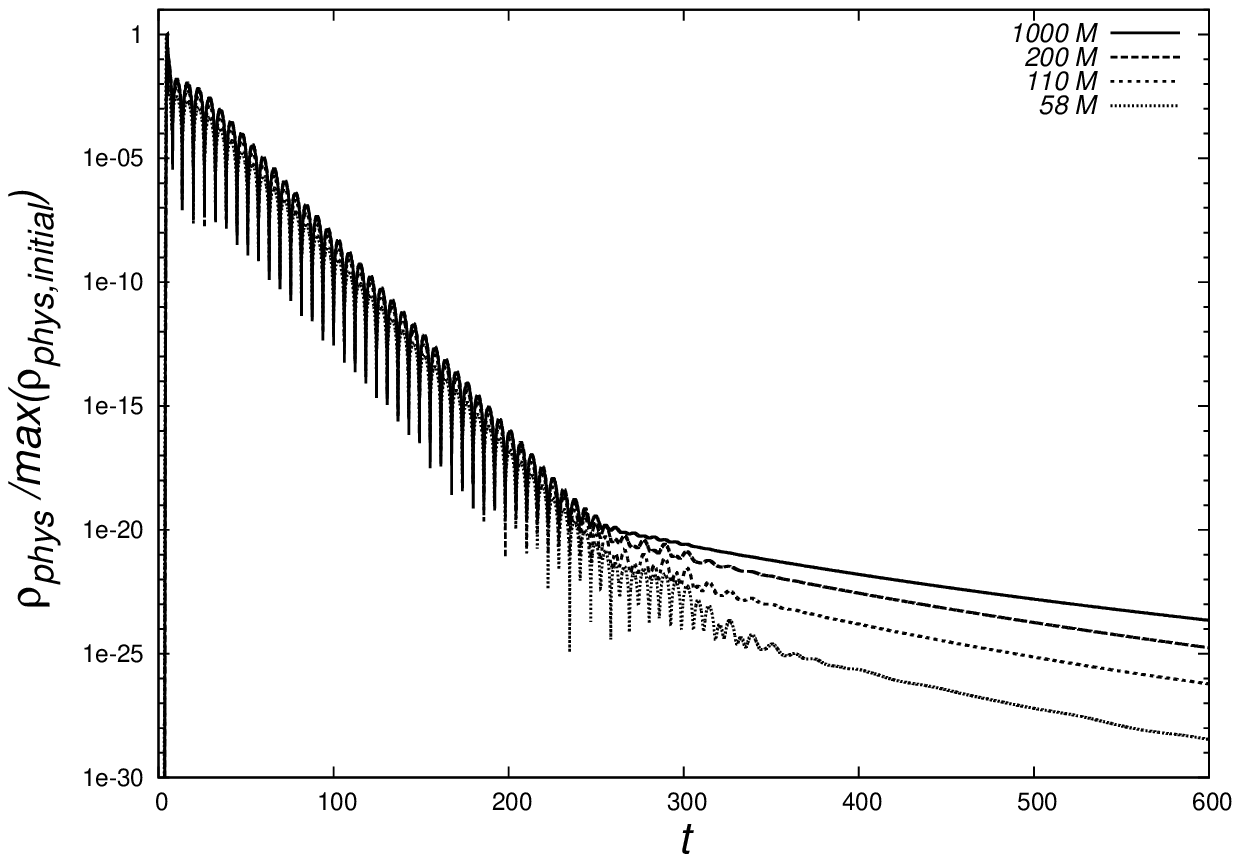}
 \hspace{3.4cm} \includegraphics[width=6cm]{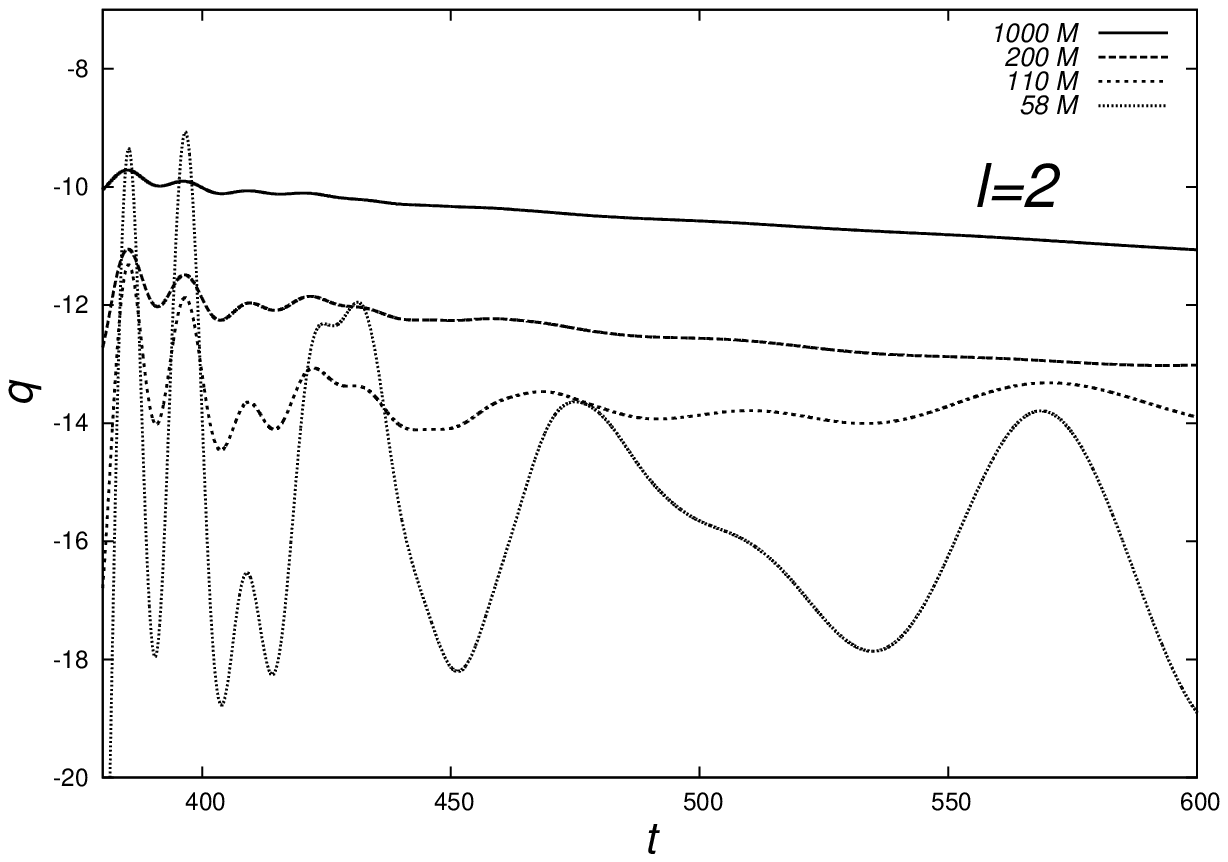}
\caption{\label{fig:rhophys} On the left we show the behavior of the physical energy density of the scalar field $\rho$ normalized with the maximum of the density at initial time, as measured by detectors located at various distances from the horizon for $l=0,1,2$ and $m_B^{2} = 0.1$. On the right we show a polynomial-like exponent that assumes the density behaves as $\sim 1/t^q$. We use these bounds in order to estimate the dilution rate of the scalar field energy density in astrophysical situations.}
\end{figure}

In Fig. \ref{fig:rhophys} we show the time dependence of the density of the scalar field $\rho$, in fact the physical density, that is, constructed with the physical and not the conformal scalar field. We again assume the density has a behavior of the type $\rho \sim 1/t^q$, in order to obtain estimates on a modest polynomial-like behavior in time. Other possibility we consider here, is the non-trivial self-interacting parameter $\lambda$. In Table \ref{table:SIP}, we present the tail decay exponent for different values of the scalar field mass $m_{B}^{2}$ and $\lambda$. In this table we also show the massless case for comparison, and we can notice that the effects on the polynomial exponent is more important when the mass is changed than when the self-interaction term is included. We consider values of $\lambda = \pm1.5$.

\begin{table}[htb]
\begin{tabular}{|c|c|c|c|c|c|}
\hline
$~~\lambda~\backslash~m_{B}^{2} ~~$ &0.0&0.1 & 0.2 \\\hline
 -1.5 &\small $p=-2.1179$ & $p=-4.0478$& $p= -6.0397$  \\\hline
  0.0 &\small $p=-2.2412$ & $p=-4.0478$& $p= -6.0398$ \\\hline
  1.5 &\small $p=-2.3952$ & $p=-4.0481$& $p= -6.0389$ \\\hline
\end{tabular}
\caption{ \label{table:SIP} In this table, we show the tail decay exponent for different values of the scalar field mass $m_{B}^{2}$ and the self-interacting parameter $\lambda$ measured by a detector located at $1000$ for $l=0$. As we can see, the effects that $\lambda$ introduces are minimal for the window of parameter here studied.  In all the cases our fittings have an uncertainty smaller than 0.2\%.
}
\end{table}

\begin{figure}[htp]
 \includegraphics[width=6cm]{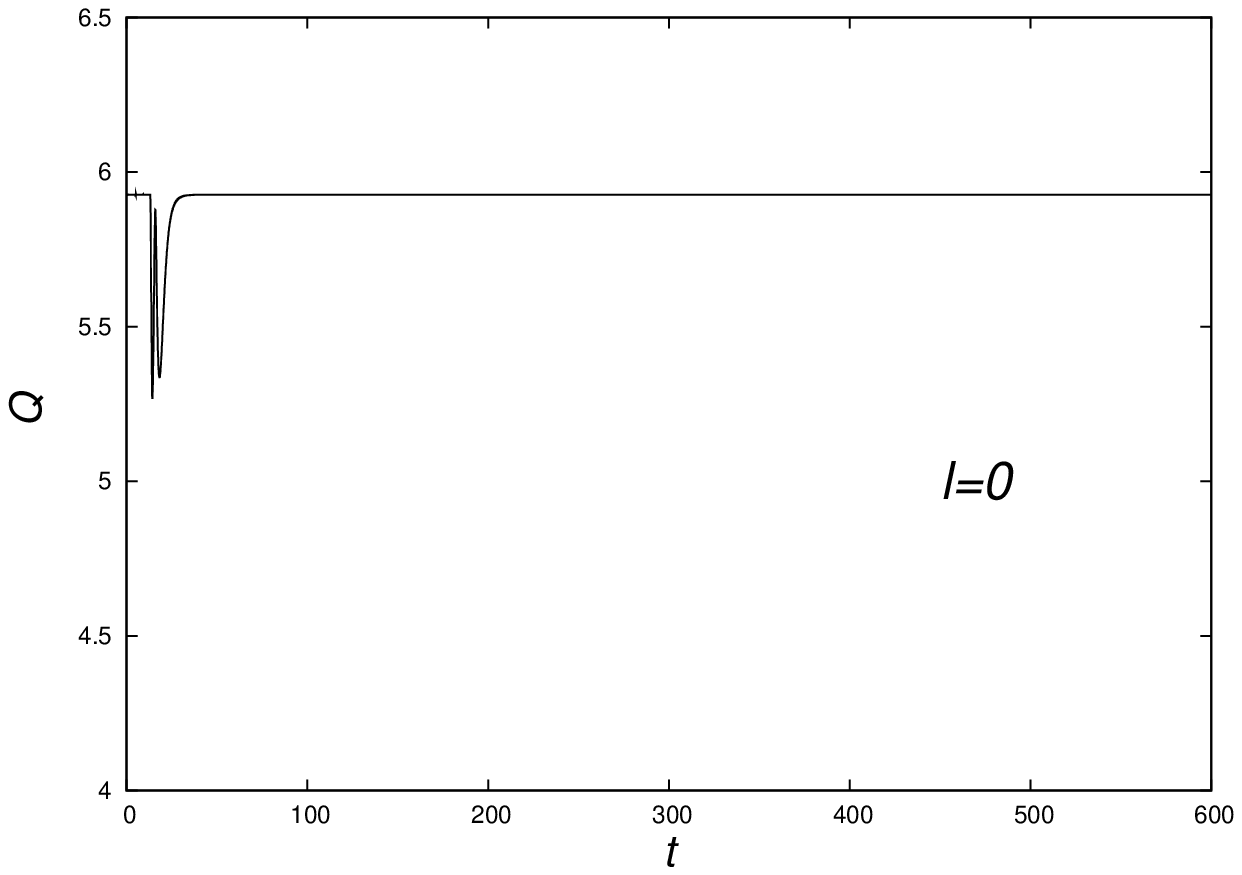}
 \includegraphics[width=6cm]{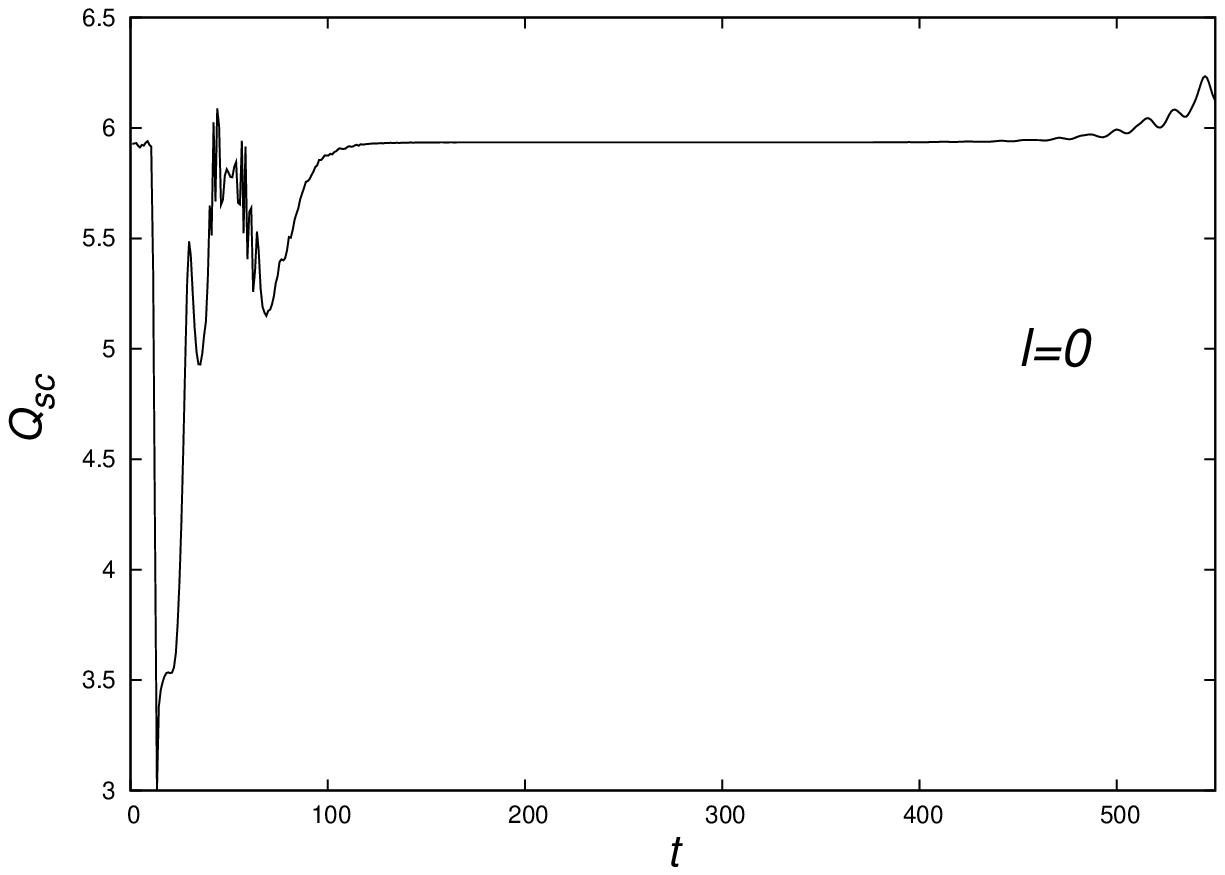}
\caption{\label{fig:conv} We show the convergence tests for one of our runs, in this case for $m_{B}^{2}=0.1$ and $l=0$ and resolutions  $\Delta x_1 = 7.6 \times 10^{-4}$, $\Delta x_2 = \Delta x_1/2$, $\Delta x_3 = \Delta x_1/4$. In the left panel we show the order of convergence $Q$ of the $L_2$ norm of the constraint ${\cal C}=\psi-\partial_r \phi$ to zero, defined by $L_2({\cal C})[using~\Delta x_2]/L_2({\cal C})[using~ \Delta x_3] = 2^Q$ because the ratio between resolutions is 2. In the right panel we show the order of self-convergence $Q_{sc}$ of $\phi$, using the $L_2$ norm of the differences between the value of the scalar field for the various resolutions.}
\end{figure}

Finally we show in Fig. \ref{fig:conv} the convergence and self-convergence of one of our simulations, showing that our algorithms work properly. The order of convergence $Q$ is consistent with our numerical implementation of sixth order in space ($Q=6$) and fourth order ($Q=4$) of the time integrator of the method of lines. Notice that the resolution in time is increased in the same proportion as the resolution in space, thus showing the convergence of the time and spatial implementation. The convergence factor appears closer to 6 than to 4 because the spatial approximation dominates over the time integration error, which is explained because after the profile of the scalar field has flattened we are evolving nearly constant functions.

{\it Astrophysical scenarios and other models.} In order to apply our numerical findings to astrophysical scenarios, we consider a supermassive black hole with mass $10^9 M_{\odot}$, and calculate the time it takes to the energy density of the scalar field to decrease five orders of magnitude for various values of $l$. We summarize our estimates in Table \ref{table:rhodecay}, where for completeness we have included the cases $m_{B}=0$ and $m_{B}^{2}=0.1$. We again construct this table considering the polynomial decay stage only.

The results indicate that the density decreases five orders of magnitude approximately every tenth of a year, independently of the initial amplitude of the scalar field, it would therefore dilute very quickly and would not survive cosmic time scales. This result is strongly related to the spherical symmetry that has been assumed, because the cross section for the accretion is maximal.

Nevertheless, more realistic models assuming spherical symmetry that consider the dark matter to be an ideal gas show different scenarios, for instance, when the gas is pressure-less it was found that within cosmic time-scales the mass of the black hole shows an exponential growth, whereas when a non-zero pressure is assumed, the pressure itself contributes to stabilize the accretion into the black hole and provides accretion mass rates more consistent with observations \cite{GuzmanLora2011}. In the present paper however, we do not measure accretion mass rates and actually focus on the considerable decrease of the scalar field density in time, because the scalar field propagates also toward {\it scri +}, whereas in the ideal gas case the particles of the fluid show speeds smaller than that of light.

Another property of the scalar field as dark matter is that a phase space analysis seems to be less practical than the solution of the Klein-Gordon equation itself; such phase space analysis would be helpful to study an adequate cross section with the black hole, and provide more realistic conditions for the scalar field to propagate, like rotation and non-zero impact parameter, compared to the spherical symmetry we present here, and also more similar to standard studies of accretion of dark matter into supermassive black holes \cite{Gilmore}.

\begin{table}[htb]
\centering
$m_{B}^{2}=0$ \\
\begin{tabular}{|c|c|c|c|c|c|}
\hline
$~~l~~$ &$~\tilde{r}~[M]~$&$~\Delta t ~[M]~$ & $\rho_{i} - \rho_{f}$ & $~\Delta t~ [years]~$ \\\hline
0 & ${\it scri +}$  & 1253& $1 \times 10^{-13} - 1 \times 10^{-18}$  &  0.195  \\
  &   54  & 661& $1 \times 10^{-13} - 1 \times 10^{-18}$  &  0.1  \\\hline
1 & ${\it scri +}$  & 1005& $1 \times 10^{-18} - 1 \times 10^{-23}$  &  0.156  \\
  &   54  & 512& $1 \times 10^{-18} - 1 \times 10^{-23}$  &  0.08  \\\hline
2 & ${\it scri +}$  & 884& $1 \times 10^{-23} - 1 \times 10^{-28}$  &  0.137  \\
  &   54  & 483 & $1 \times 10^{-25} - 1 \times 10^{-30}$  &  0.075  \\\hline
\end{tabular}
\\ $m_{B}^{2}=0.1$ \\
\begin{tabular}{|c|c|c|c|c|c|}
\hline
$~~l~~$ &$~\tilde{r}~[M]~$&$~\Delta t ~[M]~$ & $\rho_{i} - \rho_{f}$ & $~\Delta t~ [years]~$ \\\hline
0 & 1000  & 905& $1 \times 10^{-21} - 1 \times 10^{-26}$  &  0.14  \\
  &   58  & 628& $1 \times 10^{-24} - 1 \times 10^{-29}$  &  0.09  \\\hline
1 & 1000  & 680& $1 \times 10^{-22} - 1 \times 10^{-27}$  &  0.10  \\
  &   58  & 442& $1 \times 10^{-25} - 1 \times 10^{-30}$  &  0.06  \\\hline
2 & 1000  & 567& $1 \times 10^{-25} - 1 \times 10^{-30}$  &  0.08  \\
  &   58  & 319& $1 \times 10^{-28} - 1 \times 10^{-33}$  &  0.05  \\\hline
\end{tabular}
\caption{ \label{table:rhodecay} We consider the Schwarzschild solution to have the mass of a supermassive black hole with mass $M= 10^{9} M_{\odot}$. We calculate the time in years that takes the physical $\rho$ to decay its value five orders of magnitude, which we consider to be enough to decide on the possibility that the scalar field could remain around the black hole for time scales of cosmological order (gigayears).  We consider three cases $l=0,1,2$, and measure the decay time using two detectors located at different distances from the horizon ($\tilde r [M]$). We locate a time interval $\Delta t [M]$ that takes the density to decrease an amount $\rho_{i} - \rho_{f}$ and calculate it in years $\Delta t[years]$.}
\end{table}


\section{Conclusions}
\label{sec:5}

The rapid decay of the energy density of the scalar field for the case of a super-massive black hole, indicates that the scalar field cannot be maintained around the black hole during cosmological time scales in the whole space; either the scalar field is being accreted or escapes through future null infinity. The fact is that when a Schwarzschild black hole -that is asymptotically flat- is considered to be a black hole candidate, scalar fields tend to vanish from the spatial domain.

One potential ingredient that would help at maintaining massive scalar field densities during longer times is the rotation of the black hole and also of the scalar field. It would be of the major interest to use foliations that approach future null infinity and study if the same effects occur and also the study using scalar field configurations with non-zero angular momentum. In fact our results assume the maximum cross section of accretion due to the spherical symmetry we assume, which in turn works as upper bounds accretion rates in more general cases.

Yet another possibility is to consider solutions that asymptotically  containing cosmological constant, which would be appropriate if a background energy density in the universe is assumed. This would imply that black hole candidates should not be considered to be asymptotically flat. Other possibilities may include black hole candidates of a different nature like boson stars \cite{BSmimicker}.


\section*{Acknowledgments}

This research is partly supported by grants: 
CIC-UMSNH-4.9 and 
CONACyT 106466.
The runs were carried out in the IFM Cluster.


\appendix
\section{Appendix}

In this appendix we show the construction of equations (\ref{eq:1st_order}) starting from (\ref{eq:conf}).
We start with the D'Alembertian operator for a general metric written as

\begin{equation}
 \Box \phi_T = \frac{1}{\sqrt{-g}}\partial_{\mu} [\sqrt{-g}g^{\mu\nu}\partial_{\nu}\phi_T].
\end{equation}

\noindent In our spherically symmetric metric, $\sqrt{-g} = \hat\alpha \hat\gamma r^{2} \sin\theta $ is the determinant of the metric. Then (\ref{eq:conf}) reads

\begin{eqnarray}
\Box\phi_T - \frac{1}{6}R\phi_T - \left( m \Omega^{-2}\phi_T + \lambda  \phi_T^{3}\right)  &=& \frac{1}{\sqrt{-g}}\partial_{\mu}[\sqrt{-g}g^{\mu\nu}\partial_{\nu}\phi_T] \nonumber \\
&-&\frac{1}{6}R\phi_T -\left( m \Omega^{-2} \phi_T + \lambda \phi_T ^{3}\right),
\end{eqnarray}

\noindent which implies

\begin{eqnarray}
\Box\phi_T - \frac{1}{6}R\phi_T - ( m \Omega^{-2}\phi_T + \lambda  \phi_T^{3}) &=& 
\frac{1}{\hat \alpha \hat \gamma } \partial_{t}\left[ -\frac{\hat \gamma}{\hat \alpha}  \partial_{t} \phi_T +  \frac{\hat \beta \hat \gamma}{\hat\alpha} \partial_{r} \phi_T  \right] \nonumber \\
&+& \frac{1}{\hat \alpha \hat \gamma r^{2} } \partial_{r}\left[ \ r^{2} \left( \frac{\hat \gamma \hat \beta}{\hat \alpha} \right) \partial_{t} \phi_T
+ r^{2} \frac{\left( \hat \alpha^{2} -\hat \beta^{2}\hat \gamma^{2} \right) }{\hat \alpha \hat \gamma} \partial_{r} \phi_T  \right] \nonumber\\
&+& \frac{1}{ r^{2} \sin\theta} \left[\partial_{\theta} (\sin\theta \partial_{\theta}\phi_T) + \frac{1}{\sin\theta} \partial_{\varphi} (\partial_{\varphi} \phi_T) \right] \nonumber \\
&-& \frac{1}{6}R\phi_T -\left( m \Omega^{-2} \phi_T + \lambda \phi_T ^{3}\right).
\end{eqnarray}

\noindent For our first order variables

\begin{eqnarray}
 \pi &=& \frac{\hat \gamma}{\hat \alpha} \partial_{t}\phi_T - \frac{\hat \gamma}{\hat \alpha} \hat \beta \partial_{r}\phi_T, \\ \nonumber \\
 \psi &=& \partial_{r}\phi_T, \label{eq:firstorder}
\end{eqnarray}

\noindent the above equation takes the form  

\begin{eqnarray}
\partial_{t} \pi &=& \frac{1}{r^{2} } \partial_{r} \left[ \ r^{2} \left( \frac{\hat \alpha}{\hat \gamma}\psi +\hat \beta \pi \right) \right] \nonumber \\
&+& \frac{\hat \alpha \hat \gamma}{ r^{2}}\left[ \frac{1}{\sin\theta}\partial_{\theta} \left(  \sin\theta \partial_{\theta}Y_{lm}(\theta,\varphi) \right) + \frac{1}{\sin\theta^{2}} \partial_{\varphi} \left( \partial_{\
\varphi} Y_{lm}(\theta,\varphi) \right) \right]\phi(t,r) \nonumber \\
&-& \hat \alpha \hat \gamma \left[ \frac{1}{6}R\phi_T +\left( m \Omega^{-2} \phi_T + \lambda \phi_T ^{3}\right)\right],
\end{eqnarray}

\noindent where we have assumed $\phi_T(t,r,\theta,\varphi)=\phi(t,r)Y_{lm}(\theta,\varphi)$. Now, the expression between brackets in the second term is the definition of the Laplacian of spherical harmonics, which can be expresed by the following relation 

\begin{equation}
\left[ \frac{1}{\sin\theta}\partial_{\theta} \left(\sin\theta \partial_{\theta}Y_{lm}(\theta,\varphi) \right) + \frac{1}{\sin\theta^{2}} \partial_{\varphi} \left( \partial_{\varphi} Y_{lm}(\theta,\varphi) \right) \right] = -l(l+1)Y_{lm}(\theta,\varphi).
\end{equation}

\noindent Finally, we obtain the evolution equation for $\pi$ and the other two evolution equations for $\psi$ and $\phi_T$ can be obtained easily from the expressions (\ref{eq:firstorder})

\begin{eqnarray}
&&\partial_t \psi=  \partial_r\left( \frac{\hat \alpha}{\hat \gamma}\pi + \hat \beta \psi \right),\nonumber\\
&&\partial_t \pi = \frac{1}{r^2} \partial_r \left(r^2 (\hat \beta \pi + \frac{\hat \alpha}{\hat \gamma}\psi ) \right) -
\hat \alpha \hat \gamma \left( \frac{1}{6}R\phi_T + \frac{l(l+1)}{r^2}\phi_T + m_B^{2} \Omega^{-2}\phi_T + \lambda \phi_{T}^{3} \right),\nonumber\\
&&\partial_t \phi_T = \frac{\hat \alpha}{\hat \gamma} \pi + \hat \beta \psi.
\end{eqnarray}

\end{document}